\documentclass[12pt,fleqn,english]{article}
\usepackage{mathptmx}
\usepackage[T1]{fontenc}
\usepackage[latin9]{inputenc}
\usepackage{geometry}
\usepackage{color}
\usepackage{babel}
\usepackage{mathtools}
\usepackage{amsmath}
\usepackage{amssymb}
\usepackage{stmaryrd}
\usepackage{graphicx}
\usepackage{setspace}

\usepackage[authormarkup=none,final]{changes}

\definechangesauthor[name={del}, color=red]{del}
\usepackage[unicode=true,pdfusetitle,
 bookmarks=true,bookmarksnumbered=false,bookmarksopen=false,
 breaklinks=false,pdfborder={0 0 1},backref=false,colorlinks=true]
 {hyperref}
\hypersetup{
 linkcolor=blue,urlcolor={blue},filecolor={blue},citecolor={blue}}

\makeatletter
\newcommand{\lyxaddress}[1]{
	\par {\raggedright #1
	\vspace{1.4em}
	\noindent\par}
}

\usepackage{chngcntr}
\usepackage{mathtools}
\counterwithout{figure}{section}
\counterwithout{equation}{section}
\usepackage{times}
\usepackage{amsmath}
\usepackage{upgreek}

\usepackage{times}

\counterwithout{figure}{section}
\counterwithout{equation}{section}
\usepackage{changepage}

\usepackage{babel}

\counterwithout{figure}{section}
\counterwithout{equation}{section}
\usepackage{times}
\usepackage{float}
\usepackage{bm}
\usepackage[caption=false]{subfig}

\usepackage[numbers,sort&compress]{natbib} 

\date{}

\renewcommand{\fnum@figure}{Fig. \thefigure}

\@ifundefined{showcaptionsetup}{}{%
 \PassOptionsToPackage{caption=false}{subfig}}
\usepackage{subfig}
\makeatother

\begin{document}
\global\long\def\avg#1{\left\langle #1\right\rangle }%
\global\long\def\disp{u}%
\global\long\def\dens{m}%
\global\long\def\pres{F}%
\global\long\def\ini{\left(x_{0}-\Delta x\right)}%
\global\long\def\fin{\left(x_{0}+\Delta x\right)}%
\global\long\def\mid{\left(x_{0}\right)}%
\global\long\def\tey{\mid}%
\global\long\def\smidi{\left(x_{0}-1.5\frac{\Delta x}{4}\right)}%
\global\long\def\smidf{\left(x_{0}+1.5\Delta x\right)}%
\global\long\def\leftm{\left(x_{0}-0.5\Delta x\right)}%
\global\long\def\rightm{\left(x_{0}+0.5\Delta x\right)}%
 
\global\long\def\sini{\left(x_{0}-2.5\frac{\Delta x}{4}\right)}%
\global\long\def\sfin{\left(x_{0}+2.5\Delta x\right)}%

\global\long\def\sur{A}%
\global\long\def\stif{k}%
\global\long\def\t{\frac{\Delta x}{2}}%
\global\long\def\twot{\Delta x}%
\global\long\def\stress{\sigma}%
\global\long\def\dtime#1{\dot{#1}}%
\global\long\def\ddtime#1{\ddot{#1}}%
\global\long\def\dspace#1{#1'}%
\global\long\def\ddspace#1{#1''}%
\global\long\def\strain{\varepsilon}%
\global\long\def\velocity{v}%
\global\long\def\momentum{p}%
\global\long\def\ts{\frac{\Delta x}{4}}%
\global\long\def\tss{\frac{\Delta x}{8}}%
\global\long\def\effden{\tilde{\rho}}%
\global\long\def\effstif{\tilde{C}}%
\global\long\def\effdi{\tilde{A}}%
\global\long\def\effs{\hat{S}}%
\global\long\def\effsdag{\effs^{\dagger}}%
\global\long\def\teffs{\tilde{S}}%
\global\long\def\teffsdag{\teffs^{\dagger}}%
\global\long\def\el{\ell}%
\global\long\def\elea{\el_{\text{eq}}}%
\global\long\def\eleb{\elea}%
\global\long\def\ela{\elea^{a}}%
\global\long\def\elb{\elea^{b}}%
\global\long\def\auldcon{R}%
\global\long\def\masscon{\alpha}%
\global\long\def\coulomb{k_{c}}%
 
\global\long\def\zel{\el_{0}}%
\global\long\def\sel{\el^{*}}%
\global\long\def\ec{q}%
\global\long\def\dip{\Delta p_\mathrm{D}}%
\global\long\def\nochangedip{p_\mathrm{D}}%
\global\long\def\ef{E}%
\global\long\def\constant{b}%
\global\long\def\eq{\text{eq}}%
\global\long\def\dertime{\omega}%
\global\long\def\derspace{\kappa}%
\global\long\def\Amat{\mathsf{T}}%
\global\long\def\Mmat{\mathsf{M}}%
\global\long\def\Msmallmat{\hat{\Mmat}}%
\global\long\def\dvec{\mathsf{h}}%
\global\long\def\bvec{\mathsf{b}}%
\global\long\def\dipef{\dip^{\left(E\right)}}%
\global\long\def\poldens{P}%
\global\long\def\pec{\tilde{B}}%
\global\long\def\emc{\tilde{W}}%
\global\long\def\pecdag{\pec^{\dagger}}%
\global\long\def\emcdag{\emc^{\dagger}}%
\global\long\def\ed{D}%
\global\long\def\lenmod{L_{\mathrm{uc}}}%
\global\long\def\summ{\Sigma m}%
\global\long\def\sumk{\Sigma k}%
\global\long\def\wa{\dertime_{\alpha}}%
\global\long\def\wb{\dertime_{\beta}}%
\global\long\def\wc{\dertime_{\gamma}}%

\title{Discrete One-dimensional Models for the Electromomentum Coupling}
\author{Kevin Muhafra$^{1}$, Michael R.\ Haberman$^{2}$ and Gal Shmuel$^{1}$}
\maketitle

\lyxaddress{\begin{center}
$^{1}$Faculty of Mechanical Engineering, Technion--Israel Institute
of Technology, Haifa 32000, Israel\\
$^{2}$Walker Department of Mechanical Engineering, The University
of Texas at Austin, Austin, Texas 78712-1591, USA
\par\end{center}}
\begin{abstract}
Willis  dynamic homogenization theory revealed that the effective linear momentum of elastic composites is coupled to their effective strain. 
Recent generalization of Willis' dynamic homogenization theory to the case of piezoelectric composites further revealed that their effective linear momentum is also coupled to the effective electric field. Here, we introduce the simplest possible model---a one-dimensional discrete model---that exhibits this so-called electromomentum coupling in subwavelength composites. We utilize our model to elucidate the physical origins of this phenomenon, illustrate its mechanism, and identify local resonances which lead to elevated Willis- and electromomentum coupling in narrow frequency bands. The results provide intuitive guidelines for the design of this coupling in piezoelectric metamaterials. 
\end{abstract}


\section{Introduction}

Asymmetry has long been considered central in the emergence of unique physical behavior of  multiscale systems \cite{anderson1972more}. In elastodynamics,  this general notion is reinforced by the homogenization theory of \citet{Willis1985IJSS,Willis1981WM,willis1981variational,willis1997book,Willis2009,Willis2011PRSA,WILLIS2012MOM,WILLIS2012MRC}. One central discovery of his theory is that, in general, the effective constitutive relations for the linear momentum and stress fields are functions of both the strain and velocity fields. The Willis constitutive relations that arise from this dynamic homogenization procedure are nonlocal in space and time, namely, the response of a material point depends not only on the local fields at any instant in time, but also on neighboring points and their time history. The spatially local limit of the Willis equations, referred to as the Milton-Briane-Willis equations \cite{milton06cloaking,Milton2020II}, is applicable when the wavelength is much larger than the microstructure (see., e.g., the model that was first developed by \citet{Milton2007njp}), referred to as the metamaterial regime \cite{haberman2016acoustic,Cummer2016yu}, or when a single subwavelength element is analyzed\footnote{\added{By single subwavelength element, we refer to inclusion (scatterer) in a background medium   sustaining waves with length much larger than the  inclusion size ($k_0 L \ll 1$,  $k_0$ is the wavenumber in the background medium,  $L$ is the scatterer size). In this scattering regime, the response of the scatterer is well characterized by the leading order terms of the multipole expansion.}} \cite{leighton2012acoustic, leroy2002air, spratt2017radiation,lee2016origin,Muhlestein20160Prsa2}. 

The Willis couplings do not appear in the constitutive relations of the constituents, hence the resultant Willis materials are a type of metamaterial, whose behavior is fundamentally different from the behavior of their building blocks \cite{Kadi2019nrp,Christensen2015MRCComunications, craster2012acoustic,srivastava2015elastic,Simovski2009bc,lustig2019,Sridhar2018JMPS,Oh2017prapplied,SRIVASTAVA2021mom}. The surge of interest in metamaterials has resulted in a renewed interest in eponymous Willis materials, with numerous theoretical studies and experimental realizations
\cite{Meng2018prsa,nassar2015willis,Norrisrspa2011PRSA,Sieck2017prb,Srivastava2015prsa,Torrent2015PRB,Muhlestein2017nc,Melnikov2019nc}, including their application to elastic and acoustic wave control \added{for, e.g., cloaking and sound manipulation} \cite{Lawrence2020JASA,park2020as,chiang2020reconfigurable,Chen2020nc,Popa2018nc,Merkel2018prb,li2018nc,Liu2019prx,li2020prapplied}. One of the fundamental \added{theoretical} works on this topic by \citet{Sieck2017prb} addressed the physical origins of the Willis coupling, and showed that the homogenized description must include this coupling in order to be physically meaningful. \citet{Muhlestein2017nc} provided an experimental demonstration of this requirement, using the local response of a one-dimensional acoustic  element. \citet{milton06cloaking} observed that the Willis equations are analogous to the bianisotropic equations of electrodynamics, see also Refs.~\cite{Sieck2017prb,su2018prb}. 

Recently, \citet{PernasSalomon2019JMPS} generalized the theory
of Willis to account for constituents that mechanically respond to
non-mechanical stimuli, focusing on piezoelectric materials that respond to electric fields. The work demonstrated that the macroscopic linear momentum of piezoelectric composite is coupled to the electric field, and that electric displacement field is coupled to the velocity, a direct analogue to the Willis couplings. From a practical viewpoint, not only does the emergent electromomentum coupling constitute an additional degree of freedom to sense and generate elastic waves, it also opens up unique possibilities for the creation of tunable metamaterials using  external electric fields. \added{On a basic level, this occurs since the electromomentum effect, like the Willis effect,  creates a direction-dependent phase angle  \cite{rps20201wm}, which can be used for wavefront shaping. Importantly, the phase angle that the electromomentum effect generates is tunable and can be turned on and off, by changing the electric circuit conditions.} These advantages have motivated \added{further studies of the electromomentum coupling, including alternative formulations \cite{muhafra2021,lee2023arxiv}; derivation of bounds \cite{lee2022maximum,wallen2022polarizability}; optimization \cite{ZHANG2022eml,KOSTA2022ijss,HUYNH2023EML}; and application to scattering control  and cloaking \cite{Lee2023JASA,lee2022maximum}.  } 

The mechanism behind the local component of Willis-, piezoelectric-, and electromomentum effects is similar: it is the breaking some spatial symmetry in the material properties \cite{pernassalomn2020prapplied, wallen2022polarizability,lee2022maximum,Sepehrirahnama2022prl}. While continuum models provide some insight to the origins of the electromomentum effect, they yield complicated expressions for the effective coupling coefficients  which do not lend themselves to an intuitive understanding of this new material response. It is therefore advantageous to develop simpler, more intuitive models to provide a better understanding of the origins of this coupling and, in turn, enable more efficient design and fabrication of these materials. Similar models have previously been introduced to illustrate, understand, and design Willis coupling in elastic and acoustic metamaterials \cite{Milton07,Muhlestein20160Prsa2,QU2022ijms}, but have yet to be developed to analyze the electromomentum coupling. Accordingly, the objective of this work is  to provide the simplest model that illustrates the mechanism of the electromomentum coupling and elucidates its physical origins. 

To this end, we first introduce one-dimensional models for the Willis- and piezoelectric effects using systems of discrete masses, springs, and bound point charges.  We provide in Sec.\,\ref{sec: asymmetry in mass willis} our model for the Willis effect, a model which is a generalization of the model that was introduced by \citet{Muhlestein20160Prsa2}. The system we consider consists of three point masses that are connected by two linear springs. We analyze  cases where the masses and springs differ and provide expressions for the effective stiffness, mass density per unit length, and Willis coefficients. This lumped parameter model is then extended in Sec.\ \ref{sec: piezo model} to consider bound charge in order to capture the piezoelectric effect. Piezoelectricity emerges from a simple system consisting of two different masses of opposite charge that are connected by a linear spring, reminiscent of the model introduced in the classic monograph of \citet{auld1973acoustic}. In contrast with the static analysis of Auld, we consider the inertia of the masses and hence also observe local resonances and Willis effects, which may be achieved using rationally designed metamaterials that display strong Willis- and piezoelectric couplings. The piezoelectric model serves as the building block in different assemblies with which we tailor effective material properties. Specifically, we show in Sec.\ \ref{sec:EM model} that by combining two building blocks that differ by their piezoelectric coefficient, we obtain an assembly whose effective response exhibits both the Willis- and electromomentum effects. Sec.\, \ref{sec:Summary} summarizes the observations from each section of the paper and discusses implications for the design of metamaterials displaying electromomentum coupling.

\section{\label{sec: Willis models}Discrete models for the Willis coupling}
This section presents three models for Willis coupling that emerges from different types of element asymmetry. Inspired by elegantly simple model of \citet{Muhlestein20160Prsa2}, who considered a linear spring linking two different point masses, our first model uses two identical linear springs that connect three different point masses. While the model presented here exhibits the same Willis coefficient as was first shown in Ref.~\cite{Muhlestein20160Prsa2}, thanks to the middle mass it also exhibits a local resonance. Our second model consists of three identical masses that are connected by two different linear springs, a difference that also generates Willis coupling that differs from one emerging from mass asymmetry.  The last model presented in this section considers the general case where both stiffness and mass are distributed asymmetrically. \added{The results from our models are consistent with the insights from the continuum framework on Willis couplings, which show that it is a function of asymmetry in the mechanical impedance, i.e., of the mass density and elasticity.  (Explicit relations between the lumped-parameters and the continuum parameters depend on proper evaluation of continuous field relationships in the long-wavelength limit}\footnote{\added{For example, a two-layer medium with differing, but comparable, densities and elasticities can be approximated as an effective mass-spring-mass system with $m_i$ = $\rho_i L_i$ and $k = 1/\left(L_1 / M_1 + L_2 / M_2\right)$ where $m_i$ is the mass per unit area of each layer, $k$ is the spring stiffness per unit area, $L_i$ represents the layer thickness, $M_i$ is the plane wave modulus, and $i = 1,2$}.}\added{; while there are  formal means to obtain such relationships,  it is outside the scope of this work.)} 
\setlength{\belowdisplayskip}{4pt} \setlength{\belowdisplayshortskip}{4pt} \setlength{\abovedisplayskip}{4pt} \setlength{\abovedisplayshortskip}{4pt}\label{sec: asymmetry in mass willis}
\subsection{Willis coupling by mass asymmetry}
\begin{figure}
\captionsetup[subfigure]{position=top,singlelinecheck=off,justification=raggedright}\subfloat[]{{\includegraphics[width=0.53\columnwidth]{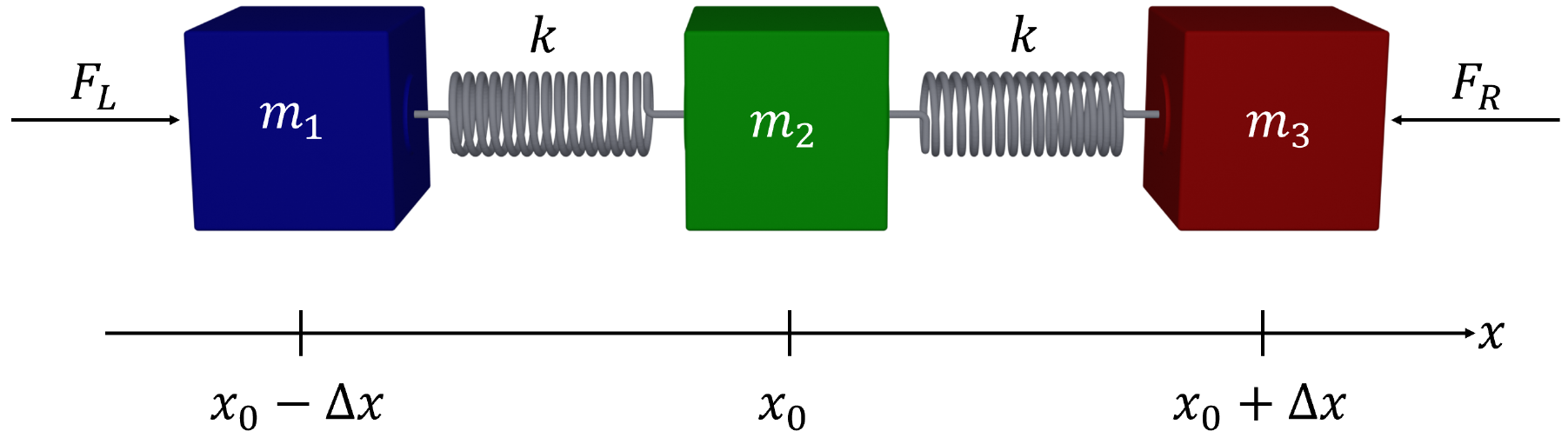}}{\small{}{}\label{fig:mass}}}\hfill{}\centering\subfloat[]{

{\includegraphics[width=0.49\columnwidth]{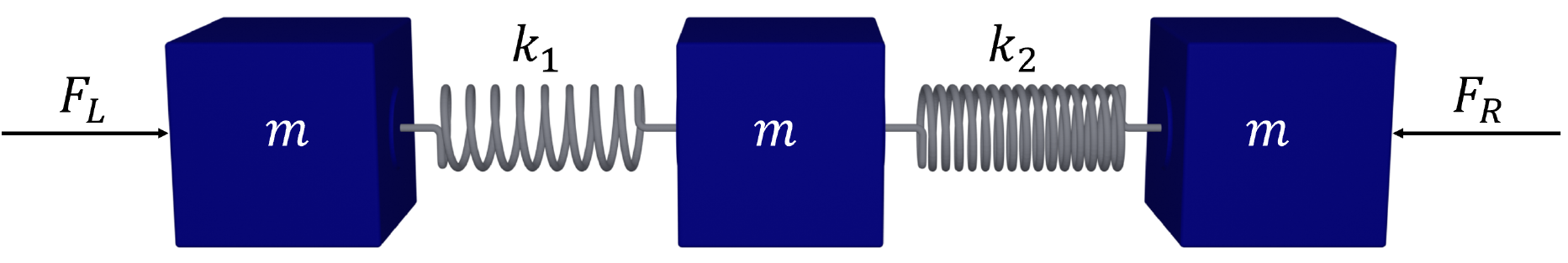}}{\small{}{}\label{fig:spring}}{\small\par}}\hfill{}\centering\subfloat[]{

{\includegraphics[width=0.47\columnwidth]{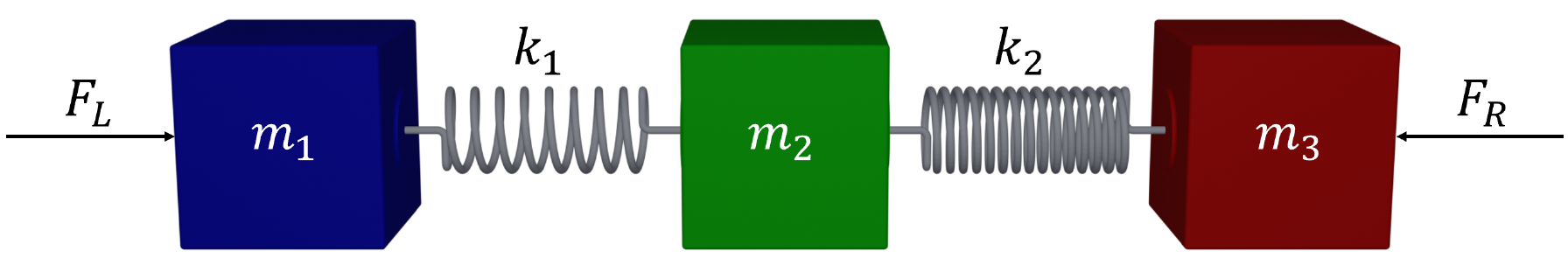}}{\small{}{}\label{fig:mass and spring}}{\small\par}}

\caption{Spring-mass models exhibiting Willis coupling owing to (a) mass asymmetry,
(b) stiffness asymmetry, and (c) combined mass-stiffness asymmetry.}
\end{figure}

Consider three different masses, namely,  $\dens_{1}$, $\dens_{2}$
and $\dens_{3}$, that are connected by linear springs of a stiffness
$\stif$, as illustrated in Fig.~\ref{fig:mass}. The system is subjected to an axial
force $\pres$ on both sides of the system; we denote by $F_\mathrm{L}$ and $F_\mathrm{R}$  the forces applied to the left and right mass, respectively, where $F_\mathrm{L}\ne F_\mathrm{R}$ in general. We assume that the time dependence of the force is harmonic of the form $e^{-i\dertime t}$. We define the mass, stiffness, and force to be normalized per unit area, thus this system represents a low-order lumped parameter model of a layered system, such as those studied by \citet{Sieck2017prb} and \citet{rps20201wm}. The resultant equations of motion of masses $\dens_{1}$, $\dens_{2}$ and $\dens_{3}$, respectively, are\begin{subequations}
\begin{align} 
\stif(\disp_\mathrm{l}-\disp_\mathrm{m})-\dens_{1}\dertime^{2}\disp_\mathrm{l}&=\pres_\mathrm{L},\label{eq:left mass M}\\
-\stif(\disp_\mathrm{l}-2\disp_\mathrm{m} + \disp_\mathrm{r})-\dens_{2}\dertime^{2}\disp_\mathrm{m}&=0,\label{eq:mid mass M}\\
\stif(\disp_\mathrm{r}-\disp_\mathrm{m})-\dens_{3}\dertime^{2}\disp_\mathrm{r}&=-\pres_\mathrm{R};\label{eq:right mass M}        
\end{align}
\end{subequations}
here, $\disp_\mathrm{l}$, $\disp_\mathrm{m}$ and $\disp_\mathrm{r}$ are the displacements of
the masses whose equilibrium position is at $x_{0}-\Delta x$, $x_{0}$,
and $x_{0}+\Delta x$, respectively, where $\twot$ is the distance between two masses at equilibrium. We
express the displacement of the middle mass as a function of the displacement of the exterior masses, such that
\begin{equation}
    \disp_{m}=\frac{\stif}{2\stif-\dens_{2}\dertime^{2}} \left(\disp_\mathrm{l}+\disp_\mathrm{r}\right) = u_\mathrm{ave}\left[1 - \left(\frac{\omega}{\omega_{\sumk\dens_{2}}}\right)^2\right]^{-1},\label{eqn:ue}
\end{equation}
where $u_\mathrm{ave} \coloneqq \left(\disp_\mathrm{l}+\disp_\mathrm{r}\right)/2$ and $\dertime_{\sumk\dens_{2}}^{2}\coloneqq2\sumk/\dens_{2}$, i.e., the sum of the stiffness divided by the mass of the middle mass. The representation on the far right-hand side of Eq.~\eqref{eqn:ue}
shows that $\disp_\mathrm{m} \rightarrow \disp_\mathrm{ave}$  for $\omega \ll \dertime_{\sumk\dens_{2}}$, where $\dertime_{\sumk\dens_{2}}$ represents a localized resonance frequency for the motion of the central mass. Substituting this expression for $\disp_\mathrm{m}$ back into Eqs.~(\ref{eq:left mass M}) and (\ref{eq:right mass M}) yields the following coupled equations of motion for $\dens_1$ and $\dens_3$
\begin{align}
-\dens_{1}\dertime^{2}\disp_\mathrm{l}+\stif\disp_\mathrm{l}-\frac{\stif}{2\stif-\dens_{2}\dertime^{2}}\left(\disp_\mathrm{l}+\disp_\mathrm{r}\right) & =\pres_\mathrm{L},\\
\dens_{3}\dertime^{2}\disp_\mathrm{r}-\stif\disp_\mathrm{r}+\frac{\stif}{2\stif-\dens_{2}\dertime^{2}}\left(\disp_\mathrm{l}+\disp_\mathrm{r}\right) & =\pres_\mathrm{R}.
\end{align}
The expressions above can be rewritten in matrix form as
\begin{equation}
\Amat\left[\begin{array}{c}
\disp_\mathrm{l}\\
\disp_\mathrm{r}
\end{array}\right]=\left[\begin{array}{c}
\pres_\mathrm{L}\\
\pres_\mathrm{R}
\end{array}\right],\text{ where }\Amat=\left[\begin{array}{cc}
-\dens_{1}\dertime^{2}+\stif-\frac{\stif^{2}}{2\stif-\dens_{2}\dertime^{2}} & -\frac{\stif^{2}}{2\stif-\dens_{2}\dertime^{2}}\\
\frac{\stif^{2}}{2\stif-\dens_{2}\dertime^{2}} & \dens_{3}\dertime^{2}-\stif+\frac{\stif^{2}}{2\stif-\dens_{2}\dertime^{2}}
\end{array}\right].\label{eq:u to N}
\end{equation}
We now limit the analysis to subwavelength microstructures which restricts the results to the metamaterial limit \cite{Sieck2017prb} by expanding each function about the center of the model, i.e., at $x=x_{0}$, and neglecting terms in the expansion of higher order than $\mathcal{O}\left(\Delta x\right)$. Using this long-wavelength approximation together with the definition of stress and the balance of linear momentum
\begin{gather}
-\pres\mid=\stress,\hspace{15pt}-\frac{\partial\pres}{\partial x}|_{x_0}\coloneqq-\dspace{\pres}\mid=\stress_{,x}=\dtime{\momentum},\label{eq: PressToStress}
\end{gather}
 yields the following coupled relationships for the stress and momentum given external forces and the strain and velocity as a function of the displacement of the left and right masses
\begin{equation}
\left[\begin{array}{c}
\stress\\
\momentum
\end{array}\right]=\Mmat^{\left(1\right)}\left[\begin{array}{c}
\pres_\mathrm{L}\\
\pres_\mathrm{R}
\end{array}\right],\quad\Mmat^{\left(1\right)}=\left[\begin{array}{cc}
-\frac{1}{2} & -\frac{1}{2}\\
-\frac{1}{2i\dertime\Delta x} & \frac{1}{2i\dertime\Delta x}
\end{array}\right],\label{eq:M1}
\end{equation}
\begin{equation}
\left[\begin{array}{c}
\strain\\
\velocity
\end{array}\right]=\Mmat^{\left(2\right)}\left[\begin{array}{c}
\disp_\mathrm{l}\\
\disp_\mathrm{r}
\end{array}\right],\quad\Mmat^{\left(2\right)}=\left[\begin{array}{cc}
-\frac{1}{2\Delta x} & \frac{1}{2\Delta x}\\
-\frac{i\dertime}{2} & -\frac{i\dertime}{2}
\end{array}\right],\label{eq:M2}
\end{equation}
where $\strain\coloneqq\partial\disp/\partial x=\dspace{\disp}$
and $\velocity\coloneqq\partial\disp/\partial t=\dtime{\disp}$ are
the strain and velocity at the center of the element, respectively. Substituting these
relations into Eq.~\eqref{eq:u to N} yields
\begin{gather}
\left[\begin{array}{c}
\stress\\
\momentum
\end{array}\right]=\Mmat^{\left(1\right)}\Amat\Mmat^{\left(2\right)^{-1}}\left[\begin{array}{c}
\strain\\
\velocity
\end{array}\right]\coloneqq\left[\begin{array}{cc}
\effstif & \teffs\\
\teffsdag & \effden
\end{array}\right]\left[\begin{array}{c}
\strain\\
\velocity
\end{array}\right],\label{eq:constitutive eq}
\end{gather}
where $\Mmat^{\left(1\right)}\Amat\Mmat^{\left(2\right)^{-1}}$ can be cast as a matrix whose entries represent the effective material properties of the subwavelength element, which can be written as
\begin{equation}
\begin{aligned}\effden= & \frac{2\stif\left(\dens_{1}+\dens_{2}+\dens_{3}\right)-\dens_{2}\dertime^{2}\left(\dens_{1}+\dens_{3}\right)}{2\Delta x\left(2\stif-\dens_{2}\dertime^{2}\right)},\\
\effstif= & \stif\Delta x-\frac{\dens_{1}+\dens_{3}}{2}\dertime^{2}\Delta x,\\
\teffs= & \teffsdag=-i\dertime\frac{\dens_{3}-\dens_{1}}{2}.
\end{aligned}
\label{eq:effprop M}
\end{equation}
We define the length of the element as $\lenmod=2\twot$ and the following frequencies of local resonance: $\dertime_{\summ}^{2}\coloneqq2\stif\summ/\left[\dens_{2}\left(\dens_{1}+\dens_{3}\right)\right]$ and $\dertime_{13}^{2}\coloneqq2\stif/\left(\dens_{1}+\dens_{3}\right)$ in addition to $\dertime_{\sumk\dens_{2}}^2$ used in Eq.~(\ref{eqn:ue}), where $\summ\coloneqq\dens_{1}+\dens_{2}+\dens_{3}$. Note that $\dertime_{\summ}^{2}$ and $\omega_{\sumk\dens_{2}}$ are related via $\dertime_{\summ}^{2} = \omega^2_{\sumk\dens_{2}} \summ/\left(m_1 + m_3\right)$, indicating that $\dertime_{\summ} > \omega_{\sumk\dens_{2}}$ for all values of $m_i$. In terms of these resonance frequencies, the effective properties can be written as
\begin{equation}
\begin{aligned}\effden= & \left(\frac{\summ}{\lenmod}\right)\frac{1-\left(\dertime/\dertime_{\summ}\right)^{2}}{1-\left(\dertime/\dertime_{\sumk\dens_{2}}\right)^{2}},\\
\effstif= & \frac{\stif}{2}\lenmod\left[1-\left(\dertime/\dertime_{13}\right)^{2}\right],\\
\teffs= & \teffsdag=-i\dertime\frac{\dens_{3}-\dens_{1}}{2} = -i\dertime\frac{\Delta m}{2},
\end{aligned}
\label{eq:newrep}
\end{equation}
where $\Delta\dens \coloneqq \dens_3 - \dens_1$ represents the mass asymmetry of the element.
 
 A few important points are observed regarding this simple model. First, we note that the frequency dependence of the dynamic effective density includes a local resonance leading to negative density for frequencies near $\omega = \omega_{\sumk\dens_{2}}$, which is when the internal mass $m_2$ oscillates out-of-phase with $m_1$ and $m_3$. Similarly, a negative effective stiffness is observed for $\omega > \omega_{13}$, where $\omega_{13}$ represents a ``breathing'' mode where $m_1$ and $m_3$ move out-of-phase with each other. Further, the Willis coefficients, $\teffs$ and $\teffsdag$ are \\(\emph{i}) equal one to another, as  they should in order to satisfy reciprocity \cite{Srivastava2015prsa,pernassalomn2020prapplied,Muhlestein20160Prsa2};  \\(\emph{ii})  linear in $\dertime$ in the low-frequency limit assumed here;  \\(\emph{iii}) of the same form as that of the mechanical impedance of a lumped mass that is equal to the difference between $\dens_1$ and $\dens_3$, hence identically zero when the element has a symmetric distribution of mass, i.e., when $\dens_1 = \dens_3$. Furthermore, if the element is inverted such that left mass equals $\dens_{3}$ and the right mass equals $\dens_{1}$, or equivalently, the coordinate system is inverted, $\teffs$ and $\teffsdag$ change their sign. This is in agreement with the fact that the Willis coupling models a direction-dependent material response related to a direction-dependent characteristic impedance, a useful property for wavefront manipulation \cite{Muhlestein2017nc,li2018nc,rps20201wm}.
Finally, representation \eqref{eq:newrep} clearly shows that the static limit yields the expected static benchmarks 

\begin{equation}
\left.\effden\right|_{\dertime\rightarrow0}=\frac{\dens_{1}+\dens_{2}+\dens_{3}}{2\Delta x} = \frac{\summ}{\lenmod},\quad\left.\effstif\right|_{\dertime\rightarrow0}=\stif\Delta x = \frac{1}{2}k\lenmod, \quad\left.\teffs\right|_{\dertime\rightarrow0}=\left.\teffsdag\right|_{\dertime\rightarrow0}=0,
\end{equation}

\noindent such that the effective mass density is the arithmetic mean of the density of the constituents, and the effective stiffness is the harmonic mean of the stiffnesses of the two springs of the element.

\subsection{Willis coupling by stiffness asymmetry}
We now modify the element to determine effective properties that result from  asymmetry in the stiffness. Accordingly, we now assume that all masses are all equal to $\dens$, while the right and left springs are represented with stiffnesses $\stif_{1}$ and $\stif_{2}$, respectively, as shown in Fig.~\ref{fig:spring}. The corresponding equations of motion of the external masses and the
middle mass, respectively, are\begin{subequations}
\begin{align} 
&-\dens\dertime^{2}\disp_\mathrm{l}+\stif_{1}(\disp_\mathrm{l}-\disp_\mathrm{m})=\pres_\mathrm{L},\label{eq:left mass S}\\
&\dens\dertime^{2}\disp_\mathrm{r}-\stif_{2}(\disp_\mathrm{r}-\disp_\mathrm{m})=\pres_\mathrm{R},\label{eq:right mass S}\\
&-\dens\dertime^{2}\disp_\mathrm{m}=\stif_{1}(\disp_\mathrm{l}-\disp_\mathrm{m})+\stif_{2}(\disp_\mathrm{r}-\disp_\mathrm{m}),\label{eq:mid mass S}
\end{align}
\end{subequations}from which we obtain 
\begin{equation}
\disp_\mathrm{m}= \left[\disp_\mathrm{l}\left(\frac{\stif_1}{\Sigma k}\right) + \disp_\mathrm{r}\left(\frac{\stif_2}{\Sigma k}\right)\right]\left[1 - \left(\frac{\dertime}{\dertime_{\sumk\dens_{2}}}\right)^2\right]^{-1},
\end{equation}
where we have defined $\sumk\coloneqq\stif_{1}+\stif_{2}$ and the local resonance $\dertime_{\sumk\dens_{2}}$ that was defined in the previous section, now for $\dens_{2}=\dens$. This expression shows that the displacement of the central mass is a weighted average of the displacements of $\disp_\mathrm{l}$ and $\disp_\mathrm{r}$ based on stiffness ratios.
Substituting this expression for $\dertime_{\sumk\dens_{2}}$ into Eqs.~(\ref{eq:left mass S}) and (\ref{eq:right mass S}) yields
\begin{equation}
\Amat\left[\begin{array}{c}
\disp_\mathrm{l}\\
\disp_\mathrm{r}
\end{array}\right]=\left[\begin{array}{c}
\pres_\mathrm{L}\\
\pres_\mathrm{R}
\end{array}\right],\text{ }\Amat=\left[\begin{array}{cc}
-\dens\dertime^{2}+\stif_{1}-\frac{\stif_{1}^{2}}{\stif_{1}+\stif_{2}-\dens\dertime^{2}} & -\left[1 - \left(\frac{\dertime}{\dertime_{\sumk\dens_{2}}}\right)^2\right]^{-1}\\
\left[1 - \left(\frac{\dertime}{\dertime_{\sumk\dens_{2}}}\right)^2\right]^{-1} & -\frac{\dens\dertime^{2}\left(\dens\dertime^{2}-2\stif_{2}\right)+\left(\stif_{2}-\dens\dertime^{2}\right)\stif_{1}}{\stif_{1}+\stif_{2}-\dens\dertime^{2}}
\end{array}\right].
\end{equation}
Following the same procedure as in the previous section and keeping frequency terms
up to order $\mathcal{O}\left(\dertime^{2}\right)$, we obtain constitutive
equations in the form of Eq.~\eqref{eq:constitutive eq}, with
\begin{equation}
\begin{aligned}\effden= & \frac{3\sumk-2\dens\dertime^{2}}{\lenmod\left(\sumk-\dens\dertime^{2}\right)}\dens,\\
\effstif= & \frac{\lenmod\left(4\stif_{1}\stif_{2}-3\dens\dertime^{2}\sumk\right)}{4\left(\sumk-\dens\dertime^{2}\right)},\\
\teffs= & \teffsdag=\frac{i\dertime\dens}{\sumk-\dens\dertime^{2}}\frac{\stif_{1}-\stif_{2}}{2},
\end{aligned}
\end{equation}
where we again note that the Willis coefficients are identically equal to
another and are only non-zero when $\stif_1 \ne \stif_2$. As such, Willis coupling vanishes when the element is symmetric, which is analogous to the fact that Willis coupling vanishes in the absence of mass asymmetry in the previous case. In terms of the local resonance frequencies
$\dertime_{3\sumk}^{2}\coloneqq3\sumk/\left(2\dens\right)$, $\wa^{2}\coloneqq4\stif_{1}/\left(3\dens\right)$
and $\wb^{2}\coloneqq4\stif_{2}/\left(3\dens\right)$, 
we may write the frequency-dependent effective properties as
\begin{equation}
\begin{aligned}\effden= & \left(\frac{3\dens}{\lenmod}\right)\frac{1-\left(\dertime/\dertime_{3\sumk}\right)^{2}}{1-\left(\dertime/\dertime_{\sumk\dens_{2}}\right)^{2}},\\
\effstif= & \lenmod\left(\frac{\stif_{1}\stif_{2}}{\sumk}\right)\frac{1-\left(\dertime/\wa\right)^{2}-\left(\dertime/\wb\right)^{2}}{1-\left(\dertime/\dertime_{\sumk\dens_{2}}\right)^{2}},\\
\teffs= & \teffsdag=\frac{1}{2}\left(\frac{\Delta k}{\sumk}\right)\frac{i\dertime\dens}{1-\left(\dertime/\dertime_{\sumk\dens_{2}}\right)^{2}},
\end{aligned}
\end{equation}
where $\Delta k \coloneqq \stif_1 - \stif_2$. Note that all effective properties are frequency dependent in the metamaterial limit, if the localized resonances occur   where the wavelength is much longer than the element. When $0<\dertime \ll \dertime_{\sumk\dens_{2}}$, the Willis coupling is linear in $\dertime$, and weighted by the ratio of the difference in stiffness and the sum of the stiffnesses. We also observe that an inverted element for which $\stif_{2}$ is the left spring and $\stif_{1}$ is the right spring, or equivalently, if the coordinate system is inverted, then $\teffs$ and $\teffsdag$ change their sign, as one would anticipate from previous research \cite{Sieck2017prb}. Notably, the  Willis coefficient exhibits a localized resonance, $\dertime_{\sumk\dens_{2}}$, unlike the Willis coefficient that emerges from   mass asymmetry. We may therefore expect that Willis coupling can be very large in a narrow band of frequencies around $\dertime_{\sumk\dens_{2}}$ when elements have an asymmetric stiffness distribution.
This representation also clearly shows that the static limit yields the following expressions properties for density, modulus, and Willis coefficients 
\begin{equation}
\left.\effden\right|_{\dertime\rightarrow0} = \frac{3\dens}{\lenmod},\quad\left.\effstif\right|_{\dertime\rightarrow0} = \frac{\stif_{1}\stif_{2}}{\sumk}\lenmod,\quad\left.\teffs\right|_{\dertime\rightarrow0}=\left.\teffsdag\right|_{\dertime\rightarrow0}=0.\label{eq:limitw}
\end{equation}
Thus, the effective mass density is the arithmetic mean of the density of the constituents, the effective stiffness is the harmonic mean of the stiffnesses of the two springs of the element, and the Willis coefficient is null.

\subsection{Willis coupling by combined stiffness and mass asymmetry}
\deleted[id=del]{In the case of}\added{Finally, we consider} a model that exhibits both mass and stiffness asymmetry as illustrated in
Fig.~\ref{fig:mass and spring}, \deleted[id=del]{the principle of superposition
delivers} \added{and repeat the same procedure as before to extract the effective properties. We find} Willis couplings that are the sum of the Willis couplings
found when only one property is asymmetric, namely,
\begin{equation}
\teffs=\teffsdag=-\frac{i\dertime}{2}\left[\Delta m+\left(\frac{\Delta k}{\sumk}\right)\frac{\dens_{2}}{1-\left(\dertime/\dertime_{\sumk\dens_{2}}\right)^{2}}\right] =\teffs_{\stif}+\teffs_{\dens},
\end{equation}
where $\teffs_{\stif}$
and $\teffs_{\dens}$ are the Willis coefficients in the asymmetric
stiffness- and asymmetric mass models, respectively. The remaining
effective properties are
\begin{gather}
\begin{aligned}\effden & =\left(\frac{\summ}{\lenmod}\right)\frac{1-\left(\dertime/\dertime_{k13}\right)^{2}}{1-\left(\dertime/\dertime_{\sumk\dens_{2}}\right)^{2}},\\
\effstif & =\lenmod\left(\frac{\stif_{1}\stif_{2}}{\sumk}\right)\frac{1-\left(\dertime/\dertime_{x\summ}\right)^{2}-\left(\dertime/\dertime_{y\summ}\right)^{2}}{1-\left(\dertime/\dertime_{\sumk\dens_{2}}\right)^{2}},
\end{aligned}
\end{gather}
where we have defined the following localized resonance frequencies: $\dertime_{k13}^{2}\coloneqq\sumk/\left[\dens_{2}\left(\dens_{1}+\dens_{3}\right)\right]$,
$\dertime_{x\summ}^{2}\coloneqq4\stif_{1}/\summ$ and $\dertime_{y\summ}^{2}\coloneqq4\stif_{2}/\summ$.
These effective properties recover the static benchmarks for $\dertime \rightarrow 0$, as
in the cases of only asymmetric mass or stiffness. These results are consistent with previous work in the literature \cite{Muhlestein2017nc, Sieck2017prb, rps20201wm}. However, due to the simplicity of the lumped parameter model considered here, these results provide insights into how local asymmetry leads to local Willis coupling. Furthermore, the relations they provide, which  include localized resonances, may be useful in designing acoustic metamaterials that have yet to be investigated.

\subsection{\added{Comparison with periodic lumped-parameter Willis model}
} 
\begin{figure}
\centerline{\includegraphics[width=0.55\columnwidth]{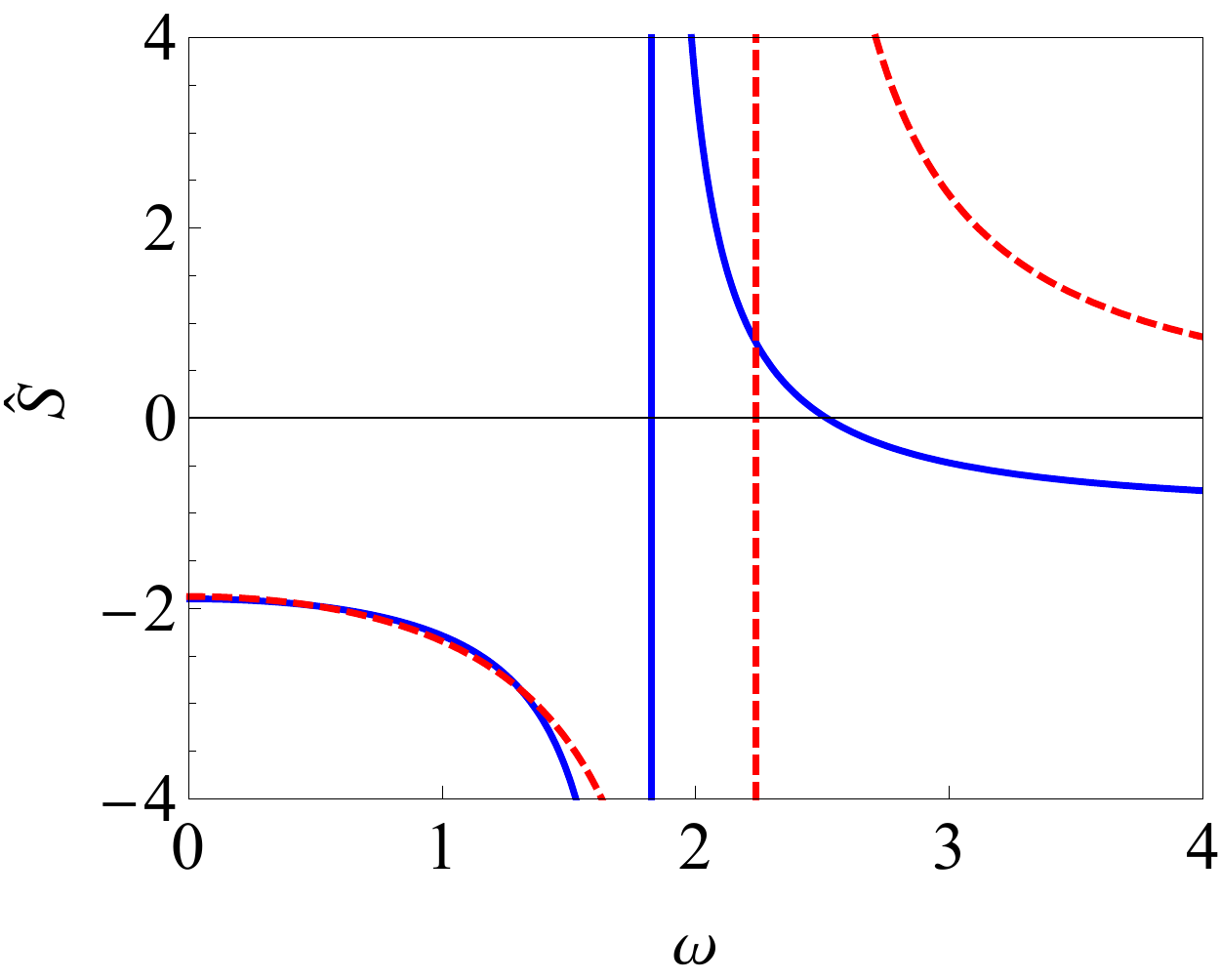}}
\caption{\label{fig:highlight}\color{black} Normalized Willis coupling $\hat{S}=i\Tilde{S}/\omega$ as function of frequency, extracted from the spatially local limit of the periodic model by \citet{nassar2015willis} (dashed red), for the representative set $k_1=1,k_2=4,m_1=1,m_2=3$, and our finite 3-mass model (solid blue), whose periodic repetition reproduces the model of  \citet{nassar2015willis} by setting $m_1=m_3=1/2$.}
\end{figure}

\added{As discussed by \citet{Simovski2010}, while effective properties that are retrieved from analysis of finite media  differ from the effective properties of bulk media, there is a qualitative relation between them, and often the former  provide  useful approximations and insights. 
It is thus advantageous to compare our finite lumped Willis model with a periodic one.  Such model was analyzed in the excellent paper of \citet{nassar2015willis}, there the authors revisited Willis dynamic homogenization method and applied it to a periodic repetition of two different masses and two springs of different stiffnesses. Their model yields spatio-temporal nonlocal effective properties, being functions of both the frequency and the Bloch wavenumber. \\
\indent We list the following similarities between the long-wavelength, low frequency limit of the Willis coupling in \citet{nassar2015willis}, and our model:\\
\indent     (\textit{i}) It is a linear function of frequency,\\
 \indent     (\textit{ii})   It is purely imaginary,\\
 \indent     (\textit{iii})    It is a function of the mass asymmetry  and stiffness-asymmetry.\\ 
 Notably, the above features are also consistent with the local limit of continuum models for the Willis coupling \cite{Sieck2017prb}.  \color{black} We illustrate these similarities in Fig.~\ref{fig:highlight}, where we numerically evaluate the normalized Willis coupling $\hat{S}\left(\omega\right)=i\Tilde{S}/\omega$, of the periodic- and finite models, for a representative example. Specifically,  local Willis coupling of the periodic model by \citet{nassar2015willis} (dashed red), is evaluated with  $k_1=1,k_2=4,m_1=1,m_2=3$, and our finite 3-mass model (solid blue) is evaluated by setting $m_1=m_3=1/2$, in order to be equivalent to a unit-cell of Nasser \emph{et al}. model. In addition to the shared features of the two models that are  listed above, we also see that both models predict local resonance, and  follow a very similar trend below that resonance frequency.\\}
\color{black}
\indent As mentioned,  while finite models  provide useful insights, they nevertheless differ from periodic models. A prominent example here is when  only one of the mechanical parameters in the periodic model of \citet{nassar2015willis} are nonuniform, in which case the model exhibits mirror symmetry and the Willis coupling vanishes. By contrast, it is clear that the mirror symmetry of the finite model is broken when either one of the two parameters is nonuniform, and indeed its Willis coupling is linear in each one of the mechanical asymmetries separately.\\
 \indent Before we proceed to develop  finite lumped-models for the piezoelectric- and electromomentum effects, we note that there are additional similarities between the effective properties of the two Willis models, e.g., both models recover the static limit of the effective mass (arithmetic mean) and effective stiffness (harmonic mean) as one would anticipate for a physically meaningful model. Collectively, the above features reinforce our approach for extracting effective properties using finite lumped-parameter models. Finally, we note the finite models that are to be derived for the piezoelectric- (Sec.~\ref{sec: piezo model}) and   electromomentum effects (Sec.~\ref{sec:EM model}), can also be extended to periodic models. In short, such extension be derived by periodically connecting the elements in Secs.~\ref{sec: piezo model}-\ref{sec:EM model},  and  analyzing one element as unit cell of a periodic medium. As such, there are additional constraints on the connection between the forces and displacements of the ends of the unit cell. The approach to extract the effective properties in the long-wavelength limit remains similar to the approach described in this work, \emph{i.e.,} obtain the equations of motion, express them in terms of the physical fields at the ends of the cell, assume that the physical fields can be represented as smooth functions, expand these functions about the center of the element in order to relate the kinematic and kinetic fields, and so on.  Bearing this in mind, we recall that the purpose of this work is to provide the simplest model that captures and elucidates the electromomentum effect and its origins at the sub-wavelength scale, a purpose that is achieved without analyzing the more complicated periodic model.

\section{\label{sec: piezo model}Discrete models of piezoelectric coupling}

\begin{figure}
\captionsetup[subfigure]{position=top,singlelinecheck=off,justification=raggedright}\centering\subfloat[]{{\includegraphics[width=0.45\columnwidth]{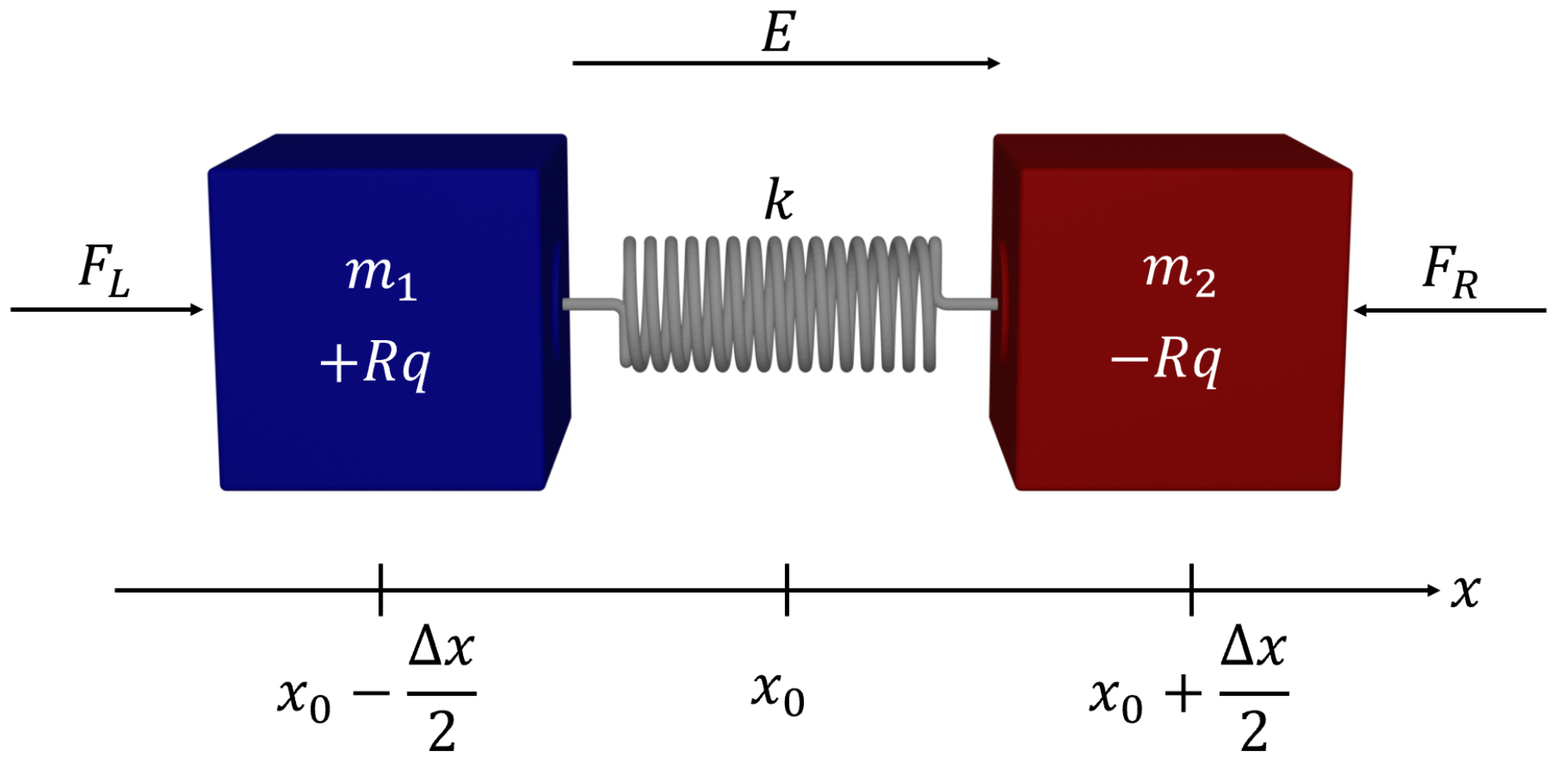}}{\small{}{}\label{fig:electric charge asymmetry}}}\hspace{0.65\columnwidth}\subfloat[]{{\includegraphics[width=0.49\columnwidth]{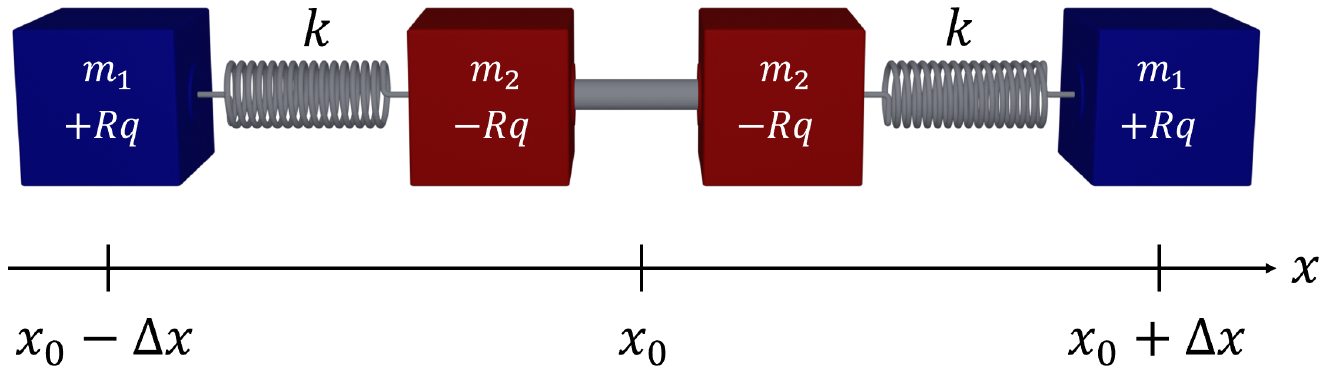}}{\small{}{}\label{fig:inverted}}}\hfill{}\subfloat[]{{\includegraphics[width=0.48\columnwidth]{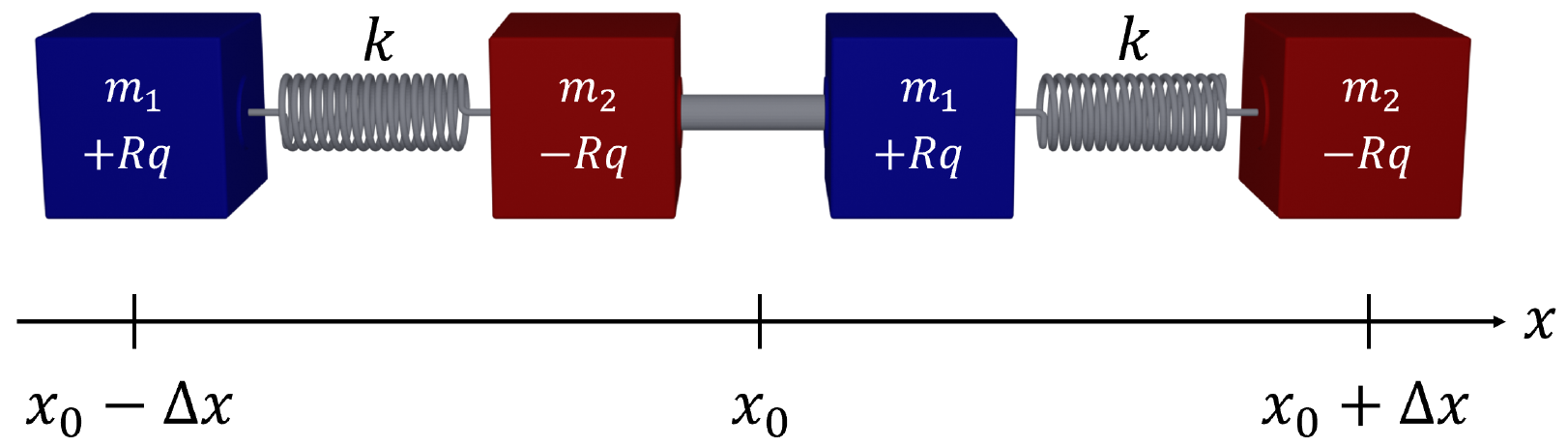}}{\small{}{}\label{fig:same}}}

\caption{Schematics of mass, spring, and point charges that demonstrate piezoelectric and Willis coupling. (a) Fundamental building block that exhibits the piezoelectric coupling. (b) A pair of building blocks from (a), attached in mirror symmetry such that there is no effective piezoelectric- or Willis coupling. (c) Assembly of these two building blocks from (a), attached with the same orientation. Since the masses and charges are distributed asymmetrically, there are effective piezoelectric and Willis couplings, however the electromomentum coupling is null since both building blocks exhibit the same piezoelectric coefficient.}
\end{figure}

Our next step is to incorporate the piezoelectric effect into the
model by extending an approach provided by  \citet{auld1973acoustic}.
We first consider two masses $\dens_{1}$ and $\dens_{2}$, connected by a linear spring of stiffness $\stif$, carrying bound electric charges of magnitude $R\ec$ and $-R\ec$, respectively, and separated by $\Delta x$, as illustrated in Fig.~\ref{fig:electric charge asymmetry}. We assume that the distance between the masses is large enough to neglect the electrical force created by the interaction between the masses, i.e., Coulomb's force, in comparison to the mechanical force, or $F_c = k_e \frac{(Rq)^2}{(\Delta x)^2} \ll \lvert k \left(\disp_\mathrm{r}-\disp_\mathrm{l}\right) \rvert $, where $k_e$ is the Coloumb constant. For the case under consideration, we must take the interaction
of bound charge with an externally applied electric field, denoted here by $\ef$,
in addition to the application of an axial pressure at the outer faces
of the model as described in the previous section. The resultant equations
of motion are 
\begin{align}
k\left(\disp_\mathrm{l}-\disp_\mathrm{r}\right)-\dens_{1}\dertime^{2}\disp_\mathrm{l}-\ef R\ec&=\pres_\mathrm{L},\label{eq:left mass Q with E}\\
-k\left(\disp_\mathrm{r}-\disp_\mathrm{l}\right)+\dens_{2}\dertime^{2}\disp_\mathrm{r}-\ef R\ec&=\pres_\mathrm{R},\label{eq:right mass Q with E}
\end{align}
which can be written in matrix form as 
\begin{equation}
\Amat\left[\begin{array}{c}
\disp_\mathrm{l}\\
\disp_\mathrm{r}
\end{array}\right]+\dvec=\left[\begin{array}{c}
\pres_\mathrm{L}\\
\pres_\mathrm{R}
\end{array}\right],\hspace{15pt} \Amat=\left[\begin{array}{cc}
-\dens_{1}\dertime^{2}+\stif & -\stif\\
\stif & \dens_{2}\dertime^{2}-\stif
\end{array}\right], \hspace{15pt} \dvec=\left[\begin{array}{c}
-R\ec\\
-R\ec
\end{array}\right]\ef.\label{eq:T=000026h two mass}
\end{equation}

We again expand each function about the center of the model at $x=x_{0}$,
and neglect terms of higher order than $\mathcal{O}\left(\Delta x\right)$.
Together with Eq.~\eqref{eq: PressToStress}, this yields the following
expressions for the local stress, momentum, strain, and velocity
\begin{equation}
\left[\begin{array}{c}
\stress\\
\momentum
\end{array}\right]=\Mmat^{\left(1\right)}\left[\begin{array}{c}
\pres_\mathrm{L}\\
\pres_\mathrm{R}
\end{array}\right],\quad\Mmat^{\left(1\right)}\ =\left[\begin{array}{cc}
-\frac{1}{2} & -\frac{1}{2}\\
-\frac{1}{i\dertime\Delta x} & \frac{1}{i\dertime\Delta x}
\end{array}\right],\label{eq:M1-2}
\end{equation}
\begin{equation}
\left[\begin{array}{c}
\strain\\
\velocity
\end{array}\right]=\Mmat^{\left(2\right)}\left[\begin{array}{c}
\disp_\mathrm{l}\\
\disp_\mathrm{r}
\end{array}\right],\quad\Mmat^{\left(2\right)}=\left[\begin{array}{cc}
-\frac{1}{\Delta x} & \frac{1}{\Delta x}\\
-\frac{i\dertime}{2} & -\frac{i\dertime}{2}
\end{array}\right].\label{eq:M2-2}
\end{equation}
Substituting these relations into Eq.~\eqref{eq:T=000026h two mass}
yields
\begin{gather}
\left[\begin{array}{c}
\stress\\
\momentum
\end{array}\right]=\Mmat^{\left(1\right)}\Amat\Mmat^{\left(2\right)^{-1}}\left[\begin{array}{c}
\strain\\
\velocity
\end{array}\right]+\Mmat^{\left(1\right)}\dvec=\left[\begin{array}{cc}
\effstif & \teffs\\
\teffsdag & \effden
\end{array}\right]\left[\begin{array}{c}
\strain\\
\velocity
\end{array}\right]-\left[\begin{array}{c}
\pecdag\\
0
\end{array}\right]\ef,\label{eq:constitutive eq-2}
\end{gather}
where the effective properties are given by 
\begin{equation}
\effden=\frac{\summ}{\lenmod},\quad\effstif=k\lenmod\left[1-\left(\dertime/\dertime_{12}\right)^{2}\right],\quad\teffs=\teffsdag=-i\dertime\frac{\Delta\dens}{2},\quad\pecdag=-\auldcon\ec;
\end{equation}
here $\dertime_{12}^{2}\coloneqq4\stif/\left(\summ\right)$,  and for this element $\lenmod=\twot$. The observations made in Sec.\ \ref{sec: Willis models} regarding the static limits of the effective density and stiffness also apply to this model, and we also observe that $\teffs$, which emerges from the asymmetric mass distribution and has the same linear dependence on frequency. In addition, we observe that the presence of asymmetric bound charge subjected to an external electric field produces an additional stress. We quantify this electric field-induced stress as $\stress_{\mathrm{E}}=\pecdag\ef$, where $\pecdag$ is the piezoelectric coefficient when the relations are cast in the stress-charge form\footnote{The piezoelectric coefficient is commonly denoted by $e$ in engineering literature and has units of $C/m^{2}$ \cite{yang1984}. However, for consistency with previous literature \cite{PernasSalomon2019JMPS,pernassalomn2020prapplied,muhafra2021,rps20201wm,KOSTA2022ijss}, we retain the notation $\pecdag$.}. Since $\pecdag$ is a property that depends on the orientation of the element, its value is flipped when the coordinate system is inverted, or when the element is flipped, such that the negative charge resides in the left mass and the positive charge resides in the right mass. We note that this directionality of the electric polarization is analogous to the directionality of the Willis coefficient in this element and in Sec.~\ref{sec: Willis models}. This analogy is the reason that mass and stiffness asymmetry of purely elastic or acoustic metamaterials can be referred to as having a \textit{Willis polarization}.

To fully characterize the electromechanical medium in the electrostatic limit, we should also consider the dielectric response of the system. To this end, we first calculate the change in the electric dipole moment due to an externally applied strain. The electric dipole moment $\nochangedip$ is defined as $\nochangedip=\Sigma\ec_{i}x_{i}$, where $\ec_{i}$ and $x_{i}$ represent the charge and position of the $i^{\mathrm{th}}$ element and $i$ is summed over all charges. When the system shown in Fig.~\ref{fig:electric charge asymmetry} is subjected to displacements on the left- and right-hand sides, the resultant  change in electric dipole moment, $\dip$, is given by
\begin{equation}
\dip=R\ec\disp_\mathrm{l}-R\ec\disp_\mathrm{r}.\label{eq:  delP def-1}
\end{equation}
This simple expression can be rewritten as
\begin{equation}
\dip=\left[\begin{array}{cc}
R\ec & -R\ec\end{array}\right]\left[\begin{array}{c}
\disp_\mathrm{l}\\
\disp_\mathrm{r}
\end{array}\right]=\left[\begin{array}{cc}
R\ec & -R\ec\end{array}\right]\Mmat^{\left(2\right)^{-1}}\left[\begin{array}{c}
\strain\\
\velocity
\end{array}\right]=-\left(R\ec\right)\lenmod\strain.
\end{equation}
The polarization density induced
by the mechanical deformation, $\poldens$, defined as the
electric dipole moment per unit volume, thus equals
\begin{equation}
\poldens=\frac{\dip}{\lenmod}=-\left(R\ec\right)\strain.\label{eq:ed}
\end{equation}
Finally, the electric displacement field in the element, $\ed$, is
the sum of this strain-induced change in electric polarization and
the electric displacement field in vacuum, i.e., 
\begin{equation}
\ed=\upepsilon_{0}\ef+\poldens=\upepsilon_{0}\ef-\left(R\ec\right)\strain,\label{eq:electric disp}
\end{equation}
where $\upepsilon_{0}$ is the permittivity of free space. Eq.\,\eqref{eq:electric disp} allows us to identify the dielectric constant $\effdi$ with $\upepsilon_{0}$ and the piezoelectric coefficient $\pec$ with $-{R\ec}$, which is equal to $\pecdag$, as it should to satisfy reciprocity \cite{pernassalomn2020prapplied}. Note that there is no relative permittivity in this case, i.e.,~ $\effdi=\upepsilon_{0}$, since our model only considers \added{two} discrete masses in free space. \added{In the more  general, the presence of a third charged mass in-between the two point masses would potentially change the electric displacement-electric field relationship, as we show in the sequel.}

In summary, Eqs.~\eqref{eq:constitutive eq-2} and \eqref{eq:electric disp}
together yield the following general constitutive relationships
\begin{equation}
\begin{aligned}\left[\begin{array}{c}
\stress\\
\ed\\
\momentum
\end{array}\right] & =\left[\begin{array}{ccc}
k\lenmod\left[1-\left(\frac{\dertime}{\dertime_{12}}\right)^2\right] & -R\ec & -i\dertime\frac{\Delta \dens}{2}\\
-R\ec & -\upepsilon_{0} & 0\\
-i\dertime\frac{\Delta \dens}{2} & 0 & \frac{\summ}{\lenmod}
\end{array}\right]\left[\begin{array}{c}
\strain\\
-\ef\\
\velocity
\end{array}\right]\\
 & \eqqcolon\left[\begin{array}{ccc}
\effstif & \pecdag & \teffs\\
\pec & -\effdi & 0\\
\teffsdag & 0 & \effden
\end{array}\right]\left[\begin{array}{c}
\strain\\
-\ef\\
\velocity
\end{array}\right].
\end{aligned}
\label{eq:effprop block1}
\end{equation}
Notably, the coupling between electrical and mechanical physics and the mechanical potential and kinetic energies is evident by the emergent non-zero off-diagonal parameters representing piezoelectricity, $\pecdag = \pec$, and Willis coupling, $\teffsdag = \teffs$, respectively. Further note that the momentum (electric displacement) and electric field and the electric displacement (velocity) are not coupled, since electromomentum coupling require asymmetry in the piezoelectric profile itself, as  we discuss in more detail in the sequel. Finally, we observe that the effective properties converge to the static benchmarks mentioned in previous sections.

We can now use this model, which demonstrates both Willis and piezoelectric effects, as the building block from which we can tailor different emergent metamaterial properties. For example, by calculating the effective properties of an assembly made of a building block and its mirror inversion, we find that the effective piezoelectric- and Willis coefficients of the system are null, as one would anticipate, since the assembly has mirror symmetry and the net polarization is null. This composition is illustrated in Fig.~\ref{fig:inverted}, where such a symmetric charge distribution (without masses) was considered by \citet{auld1973acoustic} to model a non-piezoelectric response (see figure~8.1 therein). We also note that when setting $R=1$, the model of \citet{auld1973acoustic} for a piezoelectric solid (figure~8.2 therein) is an assembly of two of our building blocks with the same orientation, as illustrated in Fig.~\ref{fig:same}. Note that Auld analyzed only the static response of the model, hence the effective properties presented there \cite{auld1973acoustic} do not include the dynamic phenomena reported here.

In view of Refs.\ \cite{PernasSalomon2019JMPS,muhafra2021,rps20201wm,KOSTA2022ijss}, which conclude that the electromomentum effect appears in composites with asymmetric piezoelectric profile\footnote{More precisely, asymmetric profile of the ratio between the piezoelectric coefficient and the dielectric coefficient.}, we expect to observe the electromomentum effect when assembling two building blocks with different piezoelectric coefficients. The analysis of such assembly is the subject of the following section.

\section{\label{sec:EM model}Discrete model of electromomentum coupling}
\begin{figure}
\captionsetup[subfigure]{position=top,singlelinecheck=off,justification=raggedright}\centering\subfloat[]{{\includegraphics[width=0.32\columnwidth]{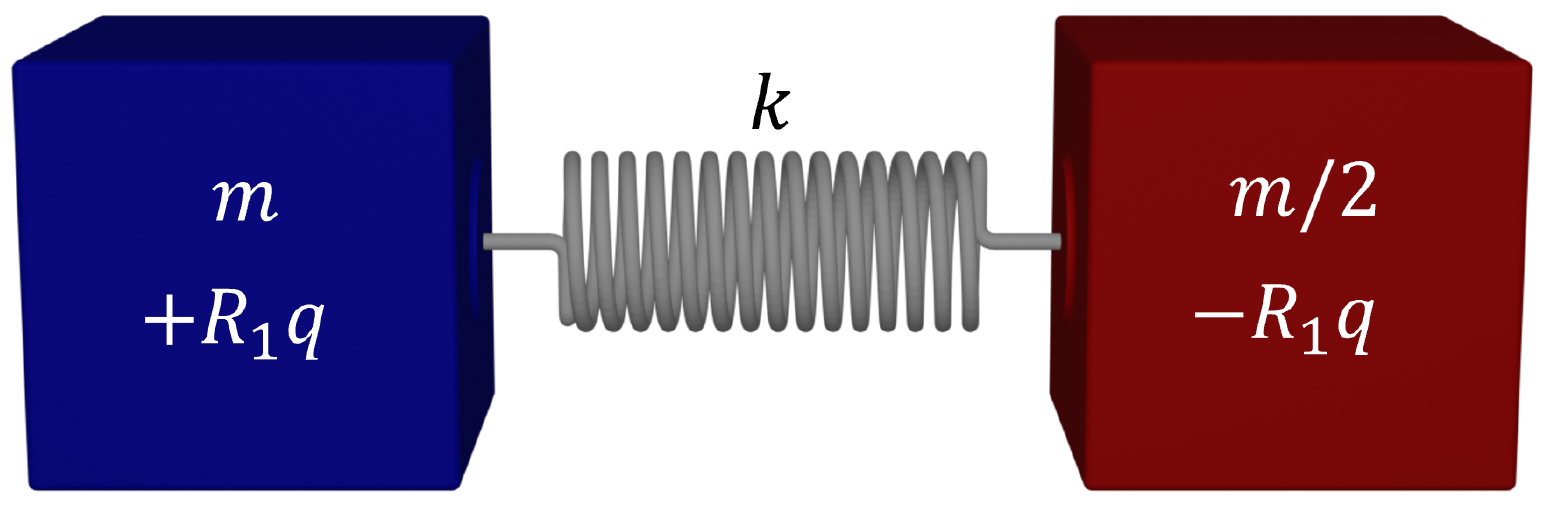}}{\small{}{}\label{fig:block1}}}\hspace{0.1\columnwidth}\subfloat[]{{\includegraphics[width=0.32\columnwidth]{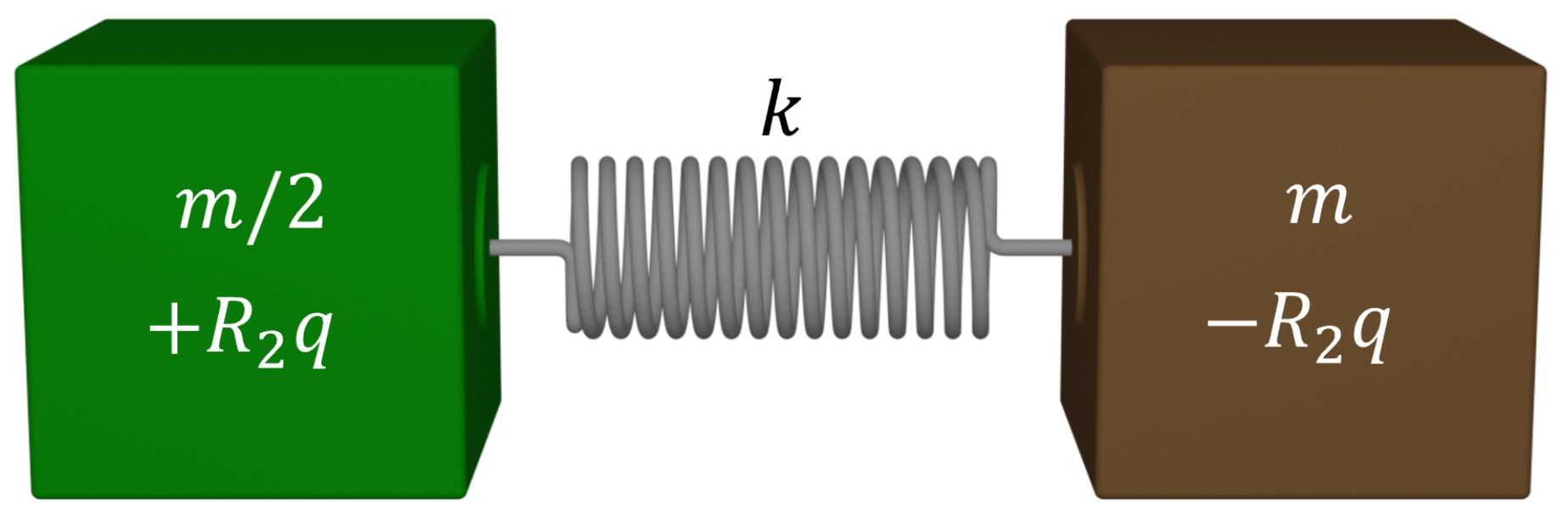}}{\small{}{}\label{fig:block2}}}\hfill{}\subfloat[]{{\includegraphics[width=0.49\columnwidth]{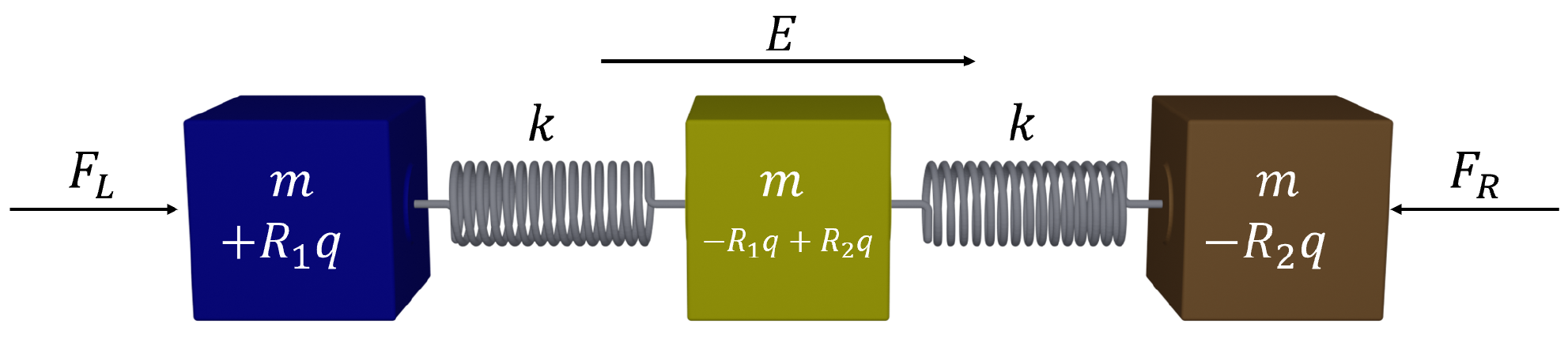}}{\small{}{}\label{fig:two blocks}}}\hspace{0.019\columnwidth}\subfloat[]{{\includegraphics[width=0.49\columnwidth]{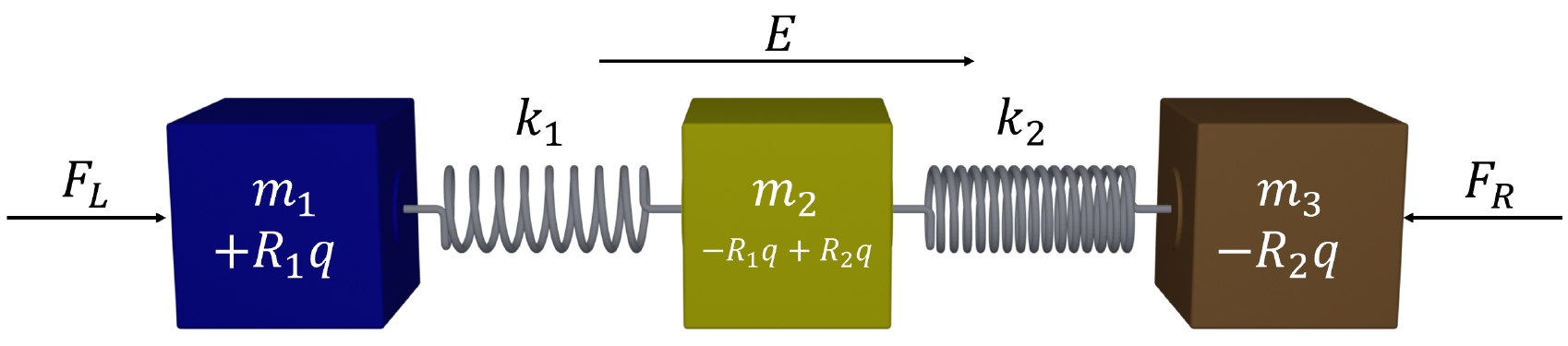}}{\small{}{}\label{fig:two blocks other spring}}}

\caption{Schematics of mass, spring, and point charges that produce piezoelectric, Willis, and electromomentum coupling. The building blocks in (a) and (b) exhibit the  piezoelectric- and Willis effects. The building block in (b) has inverse polarity in mass with respect to block (a), and therefore has the opposite Willis coefficient of block (a). It also displays a piezoelectric coefficient with a different magnitude from that of block (a) because its point charge magnitude is different. (c) Assembly of these building blocks, which exhibits the electromomentum effect. (d) Assembly of two building blocks  with two different stiffnesses and masses. Assemblies (c) and (d) exhibit both the electromomentum- and Willis couplings.}
\end{figure}

We now assemble two building blocks, the first of which is defined by $\dens_{1}=2\dens_{2}=\dens$ and $\auldcon=\auldcon_{1}$, and the second by $\dens_{2}=2\dens_{1}=\dens$ and $\auldcon=\auldcon_{2}$, as shown in Fig.~\ref{fig:block1} and Fig.~\ref{fig:block2}, respectively. Using the relations provided in Eq.~\eqref{eq:effprop block1}, we obtain the following effective properties for the building block shown in Fig.~\ref{fig:block1} 
\begin{gather}
\begin{aligned}\effden_{1} & =\frac{3\dens}{2\Delta x},\quad\effstif_{1}=\left(k+\frac{3\dens\dertime^{2}}{8}\right)\Delta x,\quad\effdi_{1}=\upepsilon_{0},\\
\teffs_{1} & =-\frac{i\dertime\dens}{4},\quad\pec_{1}=\auldcon_{1}\ec.
\end{aligned}
\label{eq:block1}
\end{gather}
The properties of the building block shown Fig.~\ref{fig:block2} are the same, except for (\emph{i}) the Willis coefficient, whose sign opposite to the sign provided in Eq.~\eqref{eq:block1}, since the masses in the second building block are flipped with respect to those in the first building block; (\emph{ii})   the piezoelectric coefficient,  which is related to the coefficient of the first building block, $\pec_{1}$,  via the ratio of their charges, i.e., $\pec_{2}=\pec_{1}\auldcon_{2}/\auldcon_{1}$. The assembly of these two building blocks as indicated in Fig.~\ref{fig:two blocks} yields a 3-mass model, since there is no spacing between the $\dens/2$ masses and they can therefore be considered as a single mass $m$. Accordingly, the left-, center-, and right-masses in the assembly are charged with $\auldcon_{1}\ec,-\auldcon_{1}\ec+\auldcon_{2}\ec$ and $-\auldcon_{2}\ec$, respectively, as shown in Fig.~\ref{fig:two blocks}. Under the application of  electric field $\ef$ and  axial force $\pres$ on the outer masses, the equations of motion are \begin{subequations}
\begin{align} 
&-\dens\dertime^{2}\disp_\mathrm{l}+\stif(\disp_\mathrm{l}-\disp_\mathrm{m})-\ef\auldcon_{1}\ec=\pres_\mathrm{L},\label{eq:left mass EM}\\
&-\dens\dertime^{2}\disp_\mathrm{m}=\stif(\disp_\mathrm{l}-\disp_\mathrm{m})+\stif(\disp_\mathrm{r}-\disp_\mathrm{m})+(\auldcon_{2}-\auldcon_{1})\ef\ec,\label{eq:mid mass EM}\\
&\dens\dertime^{2}\disp_\mathrm{r}-\stif(\disp_\mathrm{r}-\disp_\mathrm{m})-\ef\auldcon_{2}\ec=\pres_\mathrm{R},\label{eq:right mass EM}        
\end{align}
\end{subequations}
from which we have
\begin{equation}
\disp_\mathrm{m}=\frac{\stif}{2\stif-\dens\dertime^{2}}\disp_\mathrm{l}+\frac{\stif}{2\stif-\dens\dertime^{2}}\disp_\mathrm{r}+\frac{\left(\auldcon_{2}-\auldcon_{1}\right)\ef\ec}{2\stif-\dens\dertime^{2}} = \frac{1}{2}\left[\disp_\mathrm{l} + \disp_\mathrm{r} + \frac{\Delta R q}{k}E\right]\left[1 - \left(\frac{\dertime}{\dertime_{\sumk\dens_{2}}}\right)^2\right]^{-1},\label{eq:u middle q s with E}
\end{equation}
where $\Delta R \coloneqq R_2 - R_1$. We note that this expression is similar to Eq.~\eqref{eqn:ue} but with an addition contribution to the displacement, due to the interaction of the bound charge with the electric field. Substituting Eq.~(\ref{eq:u middle q s with E}) into Eqs.~(\ref{eq:left mass EM}) and (\ref{eq:right mass EM}) and rearranging terms yield
\begin{equation}
\Amat\left[\begin{array}{c}
\disp_\mathrm{l}\\
\disp_\mathrm{r}
\end{array}\right]+\dvec=\left[\begin{array}{c}
\pres_\mathrm{L}\\
\pres_\mathrm{R}
\end{array}\right],\label{eq:TandH}
\end{equation}
where 
\[
\Amat=\left[\begin{array}{cc}
-\dens\dertime^{2}+\stif-\frac{\stif^{2}}{2\stif-\dens\dertime^{2}} & -\frac{\stif^{2}}{2\stif-\dens\dertime^{2}}\\
\frac{\stif^{2}}{2\stif-\dens\dertime^{2}} & \dens\dertime^{2}-\stif+\frac{\stif^{2}}{2\stif-\dens\dertime^{2}}
\end{array}\right],\quad\dvec=-\left[\begin{array}{c}
\frac{\ec\left[\left(\stif-\dens\dertime^{2}\right)\auldcon_{1}+\stif\auldcon_{2}\right]}{2\stif-\dens\dertime^{2}}\\
\frac{\ec\left[\stif\auldcon_{1}+\left(\stif-\dens\dertime^{2}\right)\auldcon_{2}\right]}{2\stif-\dens\dertime^{2}}
\end{array}\right]\ef.
\]
By repeating the procedure introduced in Sec.~\ref{sec: piezo model} to relate $\stress$ and $\momentum$
($\strain$ and $\velocity$) to $\pres_\mathrm{L}$ and $\pres_\mathrm{R}$ ($\disp_\mathrm{l}$
and $\disp_\mathrm{r}$), we obtain
\begin{equation}
\left[\begin{array}{c}
\stress\\
\momentum
\end{array}\right]=\Mmat^{\left(1\right)}\left[\begin{array}{c}
\pres_\mathrm{L}\\
\pres_\mathrm{R}
\end{array}\right],\quad\Mmat^{\left(1\right)}=\left[\begin{array}{cc}
-\frac{1}{2} & -\frac{1}{2}\\
-\frac{1}{2i\dertime\Delta x} & \frac{1}{2i\dertime\Delta x}
\end{array}\right],\label{eq:M1-1}
\end{equation}
\begin{equation}
\left[\begin{array}{c}
\strain\\
\velocity
\end{array}\right]=\Mmat^{\left(2\right)}\left[\begin{array}{c}
\disp_\mathrm{l}\\
\disp_\mathrm{r}
\end{array}\right],\quad\Mmat^{\left(2\right)}=\left[\begin{array}{cc}
-\frac{1}{2\Delta x} & \frac{1}{2\Delta x}\\
-\frac{i\dertime}{2} & -\frac{i\dertime}{2}
\end{array}\right].\label{eq:M2-1}
\end{equation}
We then combine Eqs.\,\eqref{eq:TandH}-\eqref{eq:M2-1} to end up 
the following expressions relating the stress and momentum of the
charged mass-spring system to the externally applied strain, velocity,
and electric fields
\begin{gather}
\left[\begin{array}{c}
\stress\\
\momentum
\end{array}\right]=\Mmat^{\left(1\right)}\Amat\Mmat^{\left(2\right)^{-1}}\left[\begin{array}{c}
\strain\\
\velocity
\end{array}\right]+\Mmat^{\left(1\right)}\dvec=\left[\begin{array}{cc}
\effstif & 0\\
0 & \effden
\end{array}\right]\left[\begin{array}{c}
\strain\\
\velocity
\end{array}\right]-\left[\begin{array}{c}
\pecdag\\
\emcdag
\end{array}\right]\ef,\label{eq:constitutive eq E-1}
\end{gather}
where
\begin{gather}
\effden=\frac{3\stif\dens-\dens^{2}\dertime^{2}}{\left(2\stif-\dens\dertime^{2}\right)\Delta x},\quad\effstif=\left(k-\dens\dertime^{2}\right)\Delta x,\nonumber \\
\pecdag=-\frac{\left(\auldcon_{1}+\auldcon_{2}\right)\ec}{2}=\frac{\pecdag_{1}+\pecdag_{2}}{2},\label{eq:eff prop EM}\\
\emcdag=\frac{i\dertime\dens\left(\auldcon_{2}-\auldcon_{1}\right)\ec}{2\left(2\stif-\dens\dertime^{2}\right)\Delta x}=-\frac{i\dertime\dens}{\Delta x\left(2\stif-\dens\dertime^{2}\right)}\frac{\pecdag_{2}-\pecdag_{1}}{2}.\nonumber 
\end{gather}
Rewriting the effective properties in terms of the characteristic localized resonance frequencies $\dertime_{3km}^{2}\coloneqq3\stif/\dens$, and $\dertime_{1km}^{2}\coloneqq\stif/\dens$ yields
\begin{gather}
\effden=\frac{\summ}{\lenmod}\frac{1-\left(\dertime/\dertime_{3km}\right)^{2}}{1-\left(\dertime/\dertime_{\sumk\dens_{2}}\right)^{2}},\quad\effstif=\frac{k}{2}\lenmod\left[1-\left(\dertime/\dertime_{1km}\right)^{2}\right],\nonumber \\
\pecdag=-\frac{\left(\auldcon_{1}+\auldcon_{2}\right)\ec}{2}=\frac{\pecdag_{1}+\pecdag_{2}}{2},\label{eq:eff prop EM-1}\\
\emcdag=\frac{i\dertime\dens}{\sumk\lenmod}\left[\frac{1}{1-\left(\dertime/\dertime_{\sumk\dens_{2}}\right)^{2}}\right]\left(\pecdag_{1}-\pecdag_{2}\right)\nonumber. 
\end{gather}
The utility of the simple charged spring-mass system model is made clear through inspection of Eqs.~\eqref{eq:constitutive eq E-1}-\eqref{eq:eff prop EM-1}, which illustrate three important points about the Willis- and electromomentum couplings. First, we note that by combining two elements of opposite Willis polarization/coupling, we observe zero effective Willis coupling since the masses and springs of the assembly are distributed symmetrically. Second, the piezoelectric coefficient of the assembly is the average of the piezoelectric coefficient of its elements. If we set $\auldcon_{1}=-\auldcon_{2}$, we get $\pecdag=0$, as one would anticipate, since the charge is distributed symmetrically in that case. Importantly, the linear momentum is coupled with the electric field through $\emcdag$, which we identify as the electromomentum coupling coefficient. Its leading term in a low frequency expansion is linear in $\dertime$, and it is proportional to the asymmetry in the piezoelectric coefficient of the building blocks, i.e., $\emcdag\propto\pecdag_{2}-\pecdag_{1}$. We also identify a local resonance that can amplify $\emcdag$ when $\dertime\rightarrow\dertime_{\sumk\dens_{2}}$. This observation reinforces the continuum analysis of \citet{PernasSalomon2019JMPS}, which also identified  a resonance frequency of the electromomentum coefficient  (see figure 4 therein).

To complete the calculation of the effective relations, it is left to find expressions for the electric displacement, $\ed$, and to verify that it is indeed coupled with the velocity through some $\emc$ that is identical to $\emcdag$, a condition that results from reciprocity \cite{pernassalomn2020prapplied}. To do that, we calculate again the change in the electric dipole moment, and the corresponding polarization density $\poldens=\dip/\lenmod$, namely,
\begin{equation}
\begin{aligned}\dip & =\auldcon_{1}\ec\disp_\mathrm{l}+\left(\auldcon_{2}-\auldcon_{1}\right)\ec\disp_\mathrm{m}-\auldcon_{2}\ec\disp_\mathrm{r}=\\
 & =\frac{\Delta \auldcon^{2}\ec^{2}}{2\stif[1-\left(\dertime/\dertime_{\sumk\dens_{2}}\right)^{2}]}\ef-\frac{i\dertime\dens\Delta \auldcon\ec}{2\stif[1-\left(\dertime/\dertime_{\sumk\dens_{2}}\right)^{2}]}\velocity-\ec\Sigma \auldcon\twot\strain,
\end{aligned}
\label{eq:  delP def}
\end{equation}
where $\Sigma \auldcon=\auldcon_{1}+\auldcon_{2}$ and we have used Eqs.~\eqref{eq:u middle q s with E}, \eqref{eq:M1-1}, and \eqref{eq:M2-1} to express $\dip$ in terms of $\strain$, $\velocity$, and $\ef$. The resultant electric displacement field is then written as
\begin{equation}
\ed=\upepsilon_{0}\ef+\poldens=\left[\frac{\Delta \auldcon^{2}\ec^{2}}{2\stif\left(1-\left(\dertime/\dertime_{\sumk\dens_{2}}\right)^{2}\right)\lenmod}+\upepsilon_{0}\right]\ef-\frac{i\dertime\dens\Delta \auldcon\ec}{2\stif\left[1-\left(\dertime/\dertime_{\sumk\dens_{2}}\right)^{2}\right]\lenmod}\velocity-\frac{\Sigma \auldcon\ec}{2}\strain,\label{eq:ed-1}
\end{equation}
or $\ed=\pec\strain+\emc\velocity+\effdi\ef$, with
\begin{equation}
\begin{aligned}\pec & =\pecdag=-\frac{\Sigma \auldcon\ec}{2},\quad\emc=\emcdag=-\frac{i\dertime\dens\Delta \auldcon\ec}{2\stif\left[1-\left(\dertime/\dertime_{\sumk\dens_{2}}\right)^{2}\right]\lenmod},\\
\effdi & =\frac{\Delta \auldcon^{2}\ec^{2}}{\left(2\stif-\dens\dertime^{2}\right)\lenmod}+\upepsilon_{0}=\frac{\Delta \auldcon^{2}\ec^{2}}{2\stif\lenmod\left[1-\left(\dertime/\dertime_{\sumk\dens_{2}}\right)^{2}\right]}+\upepsilon_{0}.
\end{aligned}
\label{eq:eff prop from ed}
\end{equation}
Indeed, Eq.~\eqref{eq:eff prop from ed} shows that not only $\pec=\pecdag$,
but also that $\emc=\emcdag$, i.e., the electric displacement field
is coupled with the velocity in the same way that the linear momentum
is coupled with the electric field. Note that now the relative permittivity is not equal to zero, i.e.,~ $\effdi\ne\upepsilon_{0}$, since our model now considers three discrete masses in free space. Applying electric field will move each of the external masses freely and the middle mass will affect the polarization. Furthermore, we can see that the effective dielectric constant depends on the difference of the electric charge and on the springs' stiffness such that if $\auldcon_1=\auldcon_2$ the middle mass has no electric charge and the model will be equivalent to the model introduced in Sec.\,\ref{sec: piezo model} ($\effdi=\upepsilon_{0}$). In addition, if the stiffness will be big enough, again we will receive an equivalent model to the last model because the middle mass will not move and the ability of our model to polarize will be equivalent to two charges in free space. If we would change the electric charges of the left-, middle- and right mass to be $\auldcon_1\ec, \auldcon_2\ec, \auldcon_3\ec$ respectively, we could see that the dielectric constant depends only on the middle mass charge, i.e., $\auldcon_2\ec$. Therefore, in order to receive a non-zero relative permittivity, we must model at least three charged mass that can move in space.

Collectively, Eqs.~\eqref{eq:constitutive eq E-1}
and \eqref{eq:ed-1} establish the effective constitutive equations
of the model, which we cast in the following matrix form
\begin{equation}
\begin{aligned}\left[\begin{array}{c}
\stress\\
\ed\\
\momentum
\end{array}\right] & =\left[\begin{array}{ccc}
\stif\left[1-\left(\dertime/\dertime_{1km}\right)^{2}\right]\twot & -\frac{\Sigma \auldcon\ec}{2} & 0\\
-\frac{\Sigma \auldcon\ec}{2} & -\frac{\Delta \auldcon^{2}\ec^{2}}{2\stif\lenmod\left[1-\left(\dertime/\dertime_{\sumk\dens_{2}}\right)^{2}\right]}-\upepsilon_{0} & -\frac{i\dertime\dens\Delta \auldcon\ec}{2\stif\left[1-\left(\dertime/\dertime_{\sumk\dens_{2}}\right)^{2}\right]\lenmod}\\
0 & -\frac{i\dertime\dens\Delta \auldcon\ec}{2\stif\left[1-\left(\dertime/\dertime_{\sumk\dens_{2}}\right)^{2}\right]\lenmod} & \frac{\summ}{\lenmod}\frac{1-\left(\dertime/\dertime_{3km}\right)^{2}}{1-\left(\dertime/\dertime_{\sumk\dens_{2}}\right)^{2}}
\end{array}\right]\left[\begin{array}{c}
\strain\\
-\ef\\
\velocity
\end{array}\right]=\\
 & \eqqcolon\left[\begin{array}{ccc}
\effstif & \pecdag & 0\\
\pec & -\effdi & \emc\\
0 & \emcdag & \effden
\end{array}\right]\left[\begin{array}{c}
\strain\\
-\ef\\
\velocity
\end{array}\right].
\end{aligned}
\label{eq:matW}
\end{equation}
Note that in the absence of mechanical asymmetry, the Willis coefficients are equal to zero in Eqs.~\eqref{eq:matW},  in contrast with the analysis of the continuum model by \citet{rps20201wm}. There, the Willis coefficients are nonzero even when there is no mechanical asymmetry, provided that  the piezoelectric profile is asymmetric. We believe that  the contribution of the piezoelectric asymmetry to the Willis coefficients is absent here since we have neglected the electric force between the charges (see Sec.\,\ref{sec: piezo model}, where we assumed that $F_c\ll\lvert k \left(\disp_\mathrm{r}-\disp_\mathrm{l}\right)\rvert$). These internal electric forces 
act in effect as an additional  spring distribution, whose  profile is asymmetric when the piezoelectric profile is asymmetric, and hence would contribute to the Willis coefficients, as in the continuum model of \citet{rps20201wm}. Our assumption here that $F_c\ll\lvert k \left(\disp_\mathrm{r}-\disp_\mathrm{l}\right)\rvert$ and therefore we can neglect $F_c$ is consistent with the fact that the contribution of the asymmetry in the piezoelectric profile to the Willis coefficient in Ref.\,\cite{rps20201wm} is of order of magnitude smaller than the contribution of the mechanical asymmetry. 

We can  generalize Eq.\ \eqref{eq:matW} to a form that includes the Willis coupling by  breaking also the symmetry in the mechanical parameters, i.e., creating a difference in the springs and/or the masses (Fig.~\ref{fig:two blocks other spring}). Accordingly, we change the stiffness of the left and right springs to $\stif_{1}$ and $\stif_{2}$, respectively, and change the density of building blocks masses, such that the resultant left, middle, and right masses in the assembly are equal to $\dens_1$, $\dens_2$ and $\dens_3$, respectively. By repeating the same procedure as before, we obtain
\begin{equation}
\begin{aligned}\left[\begin{array}{c}
\stress\\
\ed\\
\momentum
\end{array}\right] &=\left[\begin{array}{ccc}
\lenmod\frac{\stif_{1}\stif_{2}}{\sumk}\frac{1-\left(\dertime/\dertime_
{\stif_{1}\stif_{2}}\right)^{2}}{1-\left(\dertime/\dertime_{\sumk\dens_{2}}\right)^{2}} & -\frac{\left[\left(2\stif_{2}-\dens_{2}\dertime^{2}\right)\auldcon_{1}+\left(2\stif_{1}-\dens_{2}\dertime^{2}\right)\auldcon_{2}\right]\ec}{2\sumk\left[1-\left(\dertime/\dertime_{\sumk\dens_{2}}\right)^{2}\right]} & -\frac{i\dertime(\Sigma\stif\Delta\dens+\Delta\stif\dens_{2})}{2\sumk\left[1-\left(\dertime/\dertime_{\sumk\dens_{2}}\right)^{2}\right]}\\
 -\frac{\left[\left(2\stif_{2}-\dens_{2}\dertime^{2}\right)\auldcon_{1}+\left(2\stif_{1}-\dens_{2}\dertime^{2}\right)\auldcon_{2}\right]\ec}{2\sumk\left[1-\left(\dertime/\dertime_{\sumk\dens_{2}}\right)^{2}\right]} & -\frac{\left(\ec\Delta R\right)^{2}}{\sumk\lenmod\left[1-\left(\dertime/\dertime_{\sumk\dens_{2}}\right)^{2}\right]}-\upepsilon_{0} & \frac{i\dertime\dens_{2}}{\lenmod\sumk}\frac{\ec\Delta R}{1-\left(\dertime/\dertime_{\sumk\dens_{2}}\right)^{2}}\\
-\frac{i\dertime(\Sigma\stif\Delta\dens+\Delta\stif\dens_{2})}{2\sumk\left[1-\left(\dertime/\dertime_{\sumk\dens_{2}}\right)^{2}\right]} & \frac{i\dertime\dens_{2}}{\lenmod\sumk}\frac{\ec\Delta R}{1-\left(\dertime/\dertime_{\sumk\dens_{2}}\right)^{2}} & \frac{\summ}{\lenmod}\frac{1-\left(\dertime/\dertime_{\sumk\summ}\right)^{2}}{1-\left(\dertime/\dertime_{\sumk\dens_{2}}\right)^{2}}
\end{array}\right]\left[\begin{array}{c}
\strain\\
-\ef\\
\velocity
\end{array}\right]\end{aligned}
\label{eq:generalprops}
\end{equation}

\noindent where $\dertime_{\stif_{1}\stif_{2}}^{2}\coloneqq4\stif_{1}\stif_{2}/\sumk\summ$
and $\dertime_{\sumk\summ}^{2}\coloneqq\Sigma\dens\Sigma\stif/\dens_{2}(\dens_{1}+\dens_{3})$. Once again, we observe that the effective properties, except $\effdi$, exhibit a resonance frequency at $\dertime_{\sumk\dens_{2}}=\sumk/\dens_{2}$. In contrast with Eq.\,\eqref{eq:matW}, we note that $\pec$ and $\teffs$ (which was null) also exhibit strong frequency-dependence due to localized resonance. Interestingly, we also observe that $\teffs$ and  $\emc$ are related via
\begin{equation}
        \emc = \frac{2q\Delta R\dens_{2}}{\lenmod(\sumk\Delta \dens+\Delta k\dens_{2})}\teffs.
\end{equation}
This implies that the relative magnitude and sign of these coefficients can be set through the ratio of the charge, stiffness, and mass contrasts, and their relative orientation in space.

\section{\label{sec:Summary}Summary}

The generalization  by \citet{PernasSalomon2019JMPS} to the homogenization scheme of \citet{Willis2011PRSA}, from elastic- to piezoelectric constitutes,  revealed that the effective linear momentum (electric displacement
field) of piezoelectric composites is coupled with the electric (velocity) field. In this work, we have developed the simplest model---a one-dimensional assembly of charged masses and springs---that exhibits these electromomentum couplings. To this end, we have first generalized the discrete models of \citet{Muhlestein20160Prsa2} and \citet{auld1973acoustic} for the Willis and piezoelectric effects, respectively. The final model, which demonstrates the electromomentum effect, comprises two elements, each of which exhibits a different piezoelectric coupling. 
\added{}
The resultant expressions for the effective properties satisfy reciprocity, recover quasistatic benchmarks, and exhibit localized resonances which may be exploited to elicit very strong narrow-band Willis- and electromomentum couplings. When the excitation frequency is much lower than the characteristic frequency, we show that the Willis- and electromomentum coefficients are linear in the frequency, and weighted by the mechanical- and electromechanical asymmetry, respectively. This conclusion reinforces and unifies previous studies that used other methods of analysis \cite{Sieck2017prb,Muhlestein20160Prsa2,PernasSalomon2019JMPS,rps20201wm,muhafra2021,KOSTA2022ijss}. While this conclusion, together with some of the other conclusions above, were reported before, here they emerge thanks to a simple model. This model allows for a more intuitive understanding and compact representation of the cross-couplings,  thereby providing a useful basis for the design of piezoelectric metamaterials.

\section*{Acknowledgments}

\added{We thank anonymous reviewers for constructive feedback that helped improve this paper}. This project was funded by the European Union (ERC, EXCEPTIONAL, Project No. 101045494); and the Israel Science Foundation, funded by the Israel Academy of Sciences and Humanities (Grant no. 2061/20). MRH acknowledges
support from the Defense Advance Research Project Agency (DARPA) and
the Army Research Office and was accomplished under Grant Number W911NF-20-1-0349. 

\bibliographystyle{unsrtnat}
\bibliography{bibtexfiletotK}

\end{document}